\documentclass[english,a4paper]{aa}
\usepackage{amssymb}
\usepackage[latin1]{inputenc}
\usepackage[T1]{fontenc}

\usepackage{rhmath}
\usepackage{rhgraphics}

\usepackage{txfonts}
\usepackage{natbib}
\bibpunct{(}{)}{;}{a}{}{,}

\usepackage{mathrsfs}
\renewcommand{\vec}[1]{\bm{#1}}
\newcommand{\mat}[1]{\tens{#1}}
\newcommand{\rot}{\nu_\textup{s}}

\usepackage{framed}
\usepackage{algpseudocode}

\begin{document}
	\title{Bayesian peak-bagging of solar-like oscillators using MCMC: \\ A comprehensive guide}
	\author{R.~Handberg\inst{1} \and T.~L.~Campante\inst{1,2}}
	\institute{Danish AsteroSeismology Centre, Department of Physics and Astronomy, Aarhus University, DK-8000 Aarhus C, Denmark \\ \email{rasmush@phys.au.dk,campante@phys.au.dk}
	\and Centro de Astrof\'isica, DFA-Faculdade de Ci\^encias, Universidade do Porto, Rua das Estrelas, 4150-762 Porto, Portugal}
	\date{Received <date> / Accepted <date>}
	\abstract
	{Asteroseismology has entered a new era with the advent of the NASA \emph{Kepler} mission. Long and continuous photometric observations of unprecedented quality are now available which have stimulated the development of a number of suites of innovative analysis tools.}
	{The power spectra of solar-like oscillations are an inexhaustible source of information on stellar structure and evolution. Robust methods are hence needed in order to infer both individual oscillation mode parameters and parameters describing non-resonant features, thus making a seismic interpretation possible.}
	{We present a comprehensive guide to the implementation of a Bayesian peak-bagging tool that employs a Markov chain Monte Carlo (MCMC). Besides making it possible to incorporate relevant prior information through Bayes' theorem, this tool also allows one to obtain the marginal probability density function for each of the fitted parameters. We apply this tool to a couple of recent asteroseismic data sets, namely, to \emph{CoRoT} observations of \object{HD 49933} and to ground-based observations made during a campaign devoted to \object{Procyon}.}
	{The developed method performs remarkably well at constraining not only in the traditional case of extracting oscillation frequencies, but also when pushing the limit where traditional methods have difficulties. Moreover it provides an rigorous way of comparing competing models, such as the ridge identifications, against the asteroseismic data.}
	{}
	
	\titlerunning{Bayesian peak-bagging of solar-like oscillators using MCMC}
	\maketitle

\section{Introduction}\label{sect:Intro}
Seismology of solar-like stars is a powerful tool that can be used to increase our understanding of stellar structure and evolution. Solar-like oscillations in main-sequence stars and subgiants have been measured thanks to data collected from ground-based high-precision spectroscopy \citep[for a review e.g.,][]{BeddingKjeldsenReviewNew} and, more recently, to photometric space-based missions such as \emph{CoRoT} \citep[e.g.,][]{Michel08}. Red giants also exhibit solar-like oscillations, although at lower frequencies, and hence require longer time series in order to resolve them \citep[e.g.,][and references therein]{Ridder09}. The launch of the NASA \emph{Kepler} mission \citep[][]{Koch10} definitely marked a milestone in the field of asteroseismology. \emph{Kepler} will particularly lead to a revolution in the seismology of solar-like oscillators, since it will increase by more than two orders of magnitude the number of stars for which high-quality observations will be available, while allowing for long-term follow-ups of a selection of those targets. The large homogeneous sample of data made available by \emph{Kepler} opens the possibility of conducting a seismic survey of the solar-like part of the colour-magnitude diagram, which researchers in the field already started naming as \emph{ensemble asteroseismology}. As of the time of writing of this article, first results arising from the \emph{Kepler} asteroseismic programme had already been made available \citep[][]{Bedding10,Chaplin10,Gilliland10,Hekker10,Stello10,JCDKjeldsenEtAl2010,Gemma}.

The rich informational content of power spectra of solar-like oscillations allows fundamental stellar properties (e.g.~mass, radius, and age) to be determined, and the internal structure to be constrained to unprecedented levels provided that individual oscillation mode parameters are measured \citep[e.g.,][]{Dalsgaard}. Furthermore, the measured stellar background signal provides us with valuable information on activity and convection. In the case of the highest signal-to-noise ratio ($S/N$) observations, for which it is possible to measure individual oscillation mode parameters, we expect asteroseismology to produce a major breakthrough on stellar structure and evolution, on topics as diverse as energy generation and transport, rotation and stellar cycles \citep[e.g.,][]{Karoff09}.

For the past few years significant work has been invested in making preparations for the mode parameter analysis of \emph{Kepler} data. This analysis involves the estimation of individual and average oscillation mode parameters, as well as estimation of parameters that describe non-resonant signatures of convection and activity. Examples include the work conducted in the framework of the AsteroFLAG consortium \citep[][]{asteroflag} and the work undertaken by the \emph{CoRoT} Data Analysis Team \citep[][]{DAT}. This consequently paved the way for the development of suites of analysis tools for application to \emph{Kepler} data \citep[][]{Birmingham,Huber,KaCa,Mathur,Mosser,ACPS}. 

In the present study we give continuity to this work by presenting a comprehensive guide to the implementation of a Bayesian peak-bagging\footnote{The term ``peak-bagging'' has become the customary name for the examination of individual oscillation peaks in the field of asteroseismology. The origin of the name is explained in \cite{Appourchaux2003B}.} tool that employs a MCMC.
These techniques derive from the tools traditionally used in helioseismology and are in many ways an extension of the Maximum Likelihood Estimation (MLE) methods.
This peak-bagging tool is to be applied to the power spectra of solar-like oscillators and used as a means to infer both individual oscillation mode parameters and parameters describing non-resonant features. Besides making it possible to incorporate relevant prior information through Bayes' theorem, this tool also allows one to obtain the marginal probability density function (PDF) for each of the model parameters (frequencies, mode heights, mode lifetimes, rotational splitting, inclination angle etc.). This is one of the main advantages of these MCMC techniques, as it not only performs well in low signal-to-noise conditions, but also provides reliable error bars on the parameters. Parameter space is sampled using a \emph{Metropolis--Hastings} algorithm featuring a built-in \emph{statistical control system} that allows to automatically set an appropriate instrumental law during the burn-in stage. Also included is \emph{parallel tempering}, which increases the mixing properties of the Markov chain.

The outline of the paper is as follows: We start in Sect.~\ref{sect:pspec} by providing an overview of the theory behind the power spectrum of solar-like oscillations, introducing the assumptions and the set of parameters needed to model the spectrum to the level of detail required by modern asteroseismic data. In Sect.~\ref{sect:Bayes} we describe the subjacent Bayesian statistical framework by highlighting the topics of parameter estimation and model selection. Section \ref{sect:MCMC} is devoted to the modus operandi of advanced Markov chain Monte Carlo methods and their implementation. In Sect.~\ref{sect:Application} we present a couple of examples where this tool has been applied to recent asteroseismic data sets, evidencing some of its capabilities and illustrating its functioning. A summary and discussion are presented in Sect.~\ref{sect:Conclusions}.

\section{The power spectrum of solar-like oscillations}\label{sect:pspec}
Solar-like oscillations or p modes (pressure playing the role of the restoring force) are global standing acoustic waves. They are characterized by being intrinsically damped while simultaneously stochastically excited by near-surface convection. Therefore, all stars cool enough to harbor an outer convective envelope -- whose locus in a H--R diagram approximately extends from the cool edge of the Cepheid instability strip up to the red giant branch -- may be expected to exhibit solar-like oscillations.

Modes of oscillation are characterized by three wave numbers: $n$, $\ell$ and $m$. The radial order $n$ characterizes the behaviour of the mode in the radial direction. The degree $\ell$ and the azimuthal order $m$ determine the spherical harmonic describing the properties of the mode as a function of colatitude and longitude. In the case of stellar observations, the associated whole-disk light integration and consequent lack of spatial resolution strongly suppress the signal from all but the modes of the lowest degree (with $\ell\!\leq\!3$). For a spherically symmetric star mode frequencies depend only on $n$ and $\ell$.

\subsection{Statistics and likelihood function of the spectrum}
Stellar p modes can be modelled as stochastically excited and intrinsically damped harmonic oscillators \citep[][]{KumarEtAl.1988}.
The frequency-power spectrum arising from such a system can in turn be modelled by a mean spectrum profile, $\mathscr{P}(\nu_j;\vec\Theta)$, described by the set of parameters $\vec\Theta$ which contain the desired physical information, multiplied by a random noise with a $\chi^2$ probability distribution with 2 degrees of freedom \citep[][]{Woodard1984,DuvallHarvey}.
This means that, at a fixed frequency bin $j$, the probability density, $f(P_j)$, that the observed power spectrum takes a particular value $P_j$, is related to the mean spectrum, $\mathscr{P}(\nu_j;\vec\Theta)$, by: 
	\begin{equation}\label{eqn:Statistics}
		f(P_j) = \frac{1}{\mathscr{P}(\nu_j;\vec\Theta)} \exp\kant*{ -\frac{P_j}{\mathscr{P}(\nu_j;\vec\Theta)} } \, .
	\end{equation}
Very often when dealing with long time series, it is customary to divide the observational data set into several independent subsets, to compute their separate spectra and to average them. In doing so one aims at decreasing the variance in the power spectrum. The average power spectrum will then obey a $\chi^2$ probability distribution with $2s$ degrees of freedom, $\chi_{2s}^2$, $s$ being the number of combined spectra \citep[][]{Appourchaux2003}:
	\begin{equation}\label{eqn:StatisticsAveraged}
		f(P_j) = \frac{s^{s-1}}{(s-1)!}\frac{P_j^{s-1}}{\mathscr{P}(\nu_j;\vec\Theta)^s} \exp\kant*{ -\frac{s \, P_j}{\mathscr{P}(\nu_j;\vec\Theta)} } \, .
	\end{equation}
Equation \eqref{eqn:StatisticsAveraged} also holds when binning the power spectrum over $s$ bins \citep[][]{Appourchaux2004}.

We would now like to specify the likelihood function, i.e., the joint PDF for the data sample $\{P_j\}$. Assuming that the frequency bins are uncorrelated, the joint PDF is simply given by the product of $f(P_j;\vec\Theta)$ over some frequency interval of interest spanned by $j$:
	\begin{equation}\label{eqn:Likelihood}
		L(\vec\Theta) = \prod_j f(P_j;\vec\Theta) \, .
	\end{equation}
Notice that we have written $f(P_j;\vec\Theta)$ to make the dependence on the parameters $\vec\Theta$ explicit. In spite of the fact that Eq.~\eqref{eqn:Likelihood} is valid for an uninterrupted data set, the same is not true when gaps are present in the time series. In that event, \citet[][]{StahnGizon} have derived an expression for the joint PDF of solar-like oscillations in complex Fourier space, in agreement with the earlier work of \citet[][]{Gabriel}. The latter PDF explicitly takes into account frequency correlations introduced by the convolution with the spectral window.

The basic idea when employing a Maximum Likelihood Estimator (MLE) is to determine estimates $\vec{\tilde\Theta}$ so as to maximize the likelihood function \citep[e.g.,][]{ToutainAppourchaux}.
Due to improved numerical stability, however, it is more convenient, in practice, to work with logarithmic probabilities:
	\begin{align}
		\mathscr{L}(\vec\Theta) &\equiv \ln L(\vec\Theta) \nonumber \\
		            &= -\sum_j \tuborg*{\ln \mathscr{P}(\nu_j;\vec\Theta) + \frac{P_j}{\mathscr{P}(\nu_j;\vec\Theta)} } \, .
	\end{align}
One therefore ends up maximizing the logarithm of the likelihood function instead:
	\begin{equation}
		\vec{\tilde\Theta} = \arg \max_{\vec\Theta} \tuborg*{ \mathscr{L}(\vec\Theta) } \, .
	\end{equation}

\subsection[Modelling the power spectrum]{Modelling the power spectrum{\footnote{To be precise, we will be modelling the power density spectrum and not the power spectrum. The former is independent of the window function and is obtained by multiplying the power spectrum by the effective length of the observational run, which can in turn be calculated as the reciprocal of the area integrated under the spectral window.}}}
The power spectrum of a single mode of oscillation is distributed around a mean profile with an exponential probability distribution according to Eq.~\eqref{eqn:Statistics}. As already mentioned, this mean profile contains the information on the physics of the mode.
In the limit of taking the ensemble average of an infinite number of realisations of the power spectrum, it can be shown \citep[][]{AndersonEtAl.1990} that the limit spectrum thus obtained follows in fact a standard Lorentzian profile near the resonance, i.e., for $\abs{\nu-\nu_0} \! \ll \! \nu_0$. A Lorentzian profile is defined as:
	\begin{equation}
		\mathscr{M}(\nu;S,\nu_0,\Gamma)=\frac{S}{1+\frac{4}{\Gamma^2}(\nu-\nu_0)^2} \, ,
	\end{equation}
where $S$ is the mode height and $\Gamma$ is the mode linewidth. $\Gamma$ is related to the mode lifetime, $\tau$, through $\Gamma\!=\!(\pi\tau)^{-1}$. In the case of solar-type stars and for low angular degree $\ell$, we can assume that $\Gamma$ is a function of frequency alone, which is supported both by observations of the Sun and by theoretical models \citep[e.g.][]{TheBook,DupretEtAl2009}.

A power spectrum of solar-like oscillations will, of course, contain a myriad of modes spanning a broad range in frequency, superimposed on a background signal of both stellar and instrumental origin. The overall limit spectrum is then given by the sum of the separate limit spectra arising from the different sources, since interference effects from beating between the modes average out in the limit.
Notice that we are assuming that a mode is uncorrelated with any other modes or with the background signal. In doing so, we neglect any eventual asymmetries of the Lorentzian profiles \citep[][]{Duvall1993,AbramsKumar1996}. Nevertheless, when dealing with long time series, such asymmetries should be included in order to avoid biases in mode frequency determination.
Furthermore, the presence of gaps and the finite length of the time series lead to a degradation of the observed power spectrum, which then results from the convolution of the true spectrum (i.e., the one that would be obtained were there no gaps) with the power spectrum of the window function (i.e., the spectral window). However, this problem is overcome by convolving the final limit spectrum with the spectral window.

Ignoring any departure from spherical symmetry, non-radial modes differing only on the azimuthal number $m$ are degenerate and their profiles will be combined into a single profile, that of the $(n,\ell)$ multiplet.
Stellar rotation removes the $(2\ell + 1)$-fold degeneracy of the frequency of oscillation of non-radial modes, thus allowing for a direct measurement of the angular velocity of the star averaged over the region probed by these modes. When the angular velocity of the star, $\Omega$, is small and in the case of rigid-body rotation, the frequency of a $(n,\ell,m)$ mode is given to first order by \citep[][]{Ledoux1951}:
	\begin{equation}
		\nu_{n\ell m} = \nu_{n\ell} + m\frac{\Omega}{2\pi}(1-C_{n \ell}) \, .
	\end{equation}
The kinematic splitting, $m\Omega/(2\pi)$, is corrected for the effect of the Coriolis force through the dimensionless quantity $C_{n\ell} \! > \! 0$. In the asymptotic regime, i.e., for high-order, low-degree p modes, rotational splitting is dominated by advection and the splitting between adjacent modes within a multiplet is $\rot \!\! \simeq \!\! \Omega/(2\pi)$. Second-order rotational effects are related to the distortion of the equilibrium structure of the star caused by centrifugal forces. Although negligible in the Sun, these effects are significant for faster solar-type rotators where these effects can cause non-negligible biases on frequency determinations \citep[e.g.,][]{Ballot2010}. Large-scale magnetic fields may also introduce further corrections to the oscillation frequencies.

\begin{figure}[tb]
	\centering
	\includegraphics[width=0.48\textwidth]{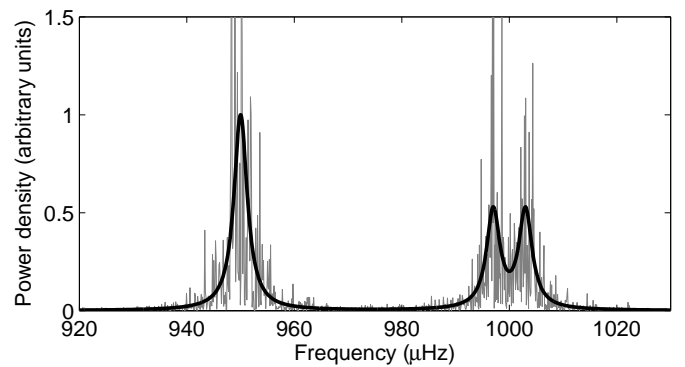}
	\caption{Artificial power density spectrum of a $\ell\!=\!0$ singlet and a $\ell\!=\!1$ multiplet. One has $\rot\!=\!\Gamma\!=\!\SI{3}{\uHz}$. Notice how the $\ell\!=\!1$ multiplet splits into its $m$ components. The power spectrum (grey) is distributed around a mean spectrum (black) with an exponential probability distribution.}
	\label{fig:Multiplet}
\end{figure}

Assuming energy equipartition between the components of a multiplet, we define the following symmetric profile for a $(n,\ell)$ multiplet:
	\begin{equation}\label{eqn:Multiplet}
			\mathscr{M}_{n\ell}(\nu;S_{n\ell},\nu_{n\ell},\Gamma_{n\ell},\rot,i) = \sum_{m=-\ell}^{\ell} \mathscr{E}_{\ell m}(i) \, \mathscr{M}(\nu;S_{n\ell},\nu_{n\ell}+m\rot,\Gamma_{n\ell}) \, ,
	\end{equation}
where $\mathscr{E}_{\ell m}(i)$ represents mode visibility within a multiplet and $i$ is the inclination angle between the direction of the stellar rotation axis and the line of sight.
The overall profile of a multiplet thus consists of the sum of $2\ell+1$ Lorentzian profiles regularly spaced in frequency and scaled in height according to the $\mathscr{E}_{\ell m}(i)$ factors (see Fig.~\ref{fig:Multiplet}), which in turn are given by \citep[][]{GizonSolanki}:     
	\begin{equation}\label{eqn:Elm}
		\mathscr{E}_{\ell m}(i) = \frac{(\ell-|m|)!}{(\ell+|m|)!} \kant*{ P_\ell^{\abs{m}}(\cos i) }^2 \, ,
	\end{equation}
where $P_\ell^m(x)$ are the associated Legendre functions. Notice that $\sum_m \mathscr{E}_{\ell m}(i) \! = \! 1$, meaning that $\mathscr{E}_{\ell m}(i)$ represents the relative power contained in a mode within a multiplet. 

Since we are primarily interested in performing a so-called \emph{global fit} \citep[e.g.,][]{Appourchaux08} to the observed power spectrum, whereby several radial orders are fitted simultaneously within a broad frequency range, we end up modelling the mean acoustic spectrum according to the following general relation:
	\begin{equation}\label{eqn:LimitSpectrum}
		\mathscr{P}(\nu;\vec\Theta) = \sum_{n=n_0}^{n_\textup{max}} \sum_{\ell=0}^{\ell_\textup{max}} \sum_{m=-\ell}^\ell \frac{ \mathscr{E}_{\ell m}(i) \, S_{n\ell}}{ 1 + \frac{4}{\Gamma_{n\ell}^2} (\nu - \nu_{n\ell} - m \rot)^2 }  + N(\nu) \, ,
	\end{equation}
where we have also included a profile describing the background signal, $N(\nu)$. Granulation, faculae and active regions might contribute to the stellar background signal, which is commonly modelled as a sum of power laws describing these physical phenomena \citep[][]{Harvey,AigrainEtAl.2004}:
	\begin{equation}
		N(\nu) = \sum_{k=1}^{k_\textup{max}}\frac{4 A_k^2 B_k}{1+(2\pi B_k \nu)^{C_k}} + N \, ,
	\label{eqn:Harvey}	
	\end{equation}
$\{A_k\}$ and $\{B_k\}$ being, respectively, the corresponding amplitudes and characteristic time-scales, whereas the $\{C_k\}$ are the slopes of each of the individual power laws. A flat component, $N$, is needed in order to model the photon shot noise. Equation \eqref{eqn:Harvey} might just well incorporate any instrumental background signal. We refer to $S_{n\ell}/N(\nu_{n\ell})$ as the signal-to-noise ratio (in power) of the multiplet $(n,\ell)$. 

Once again assuming energy equipartition between the different components of a multiplet, their heights can be expressed as:
	\begin{equation}\label{eqn:Ballot}
		S_{n \ell m}=\mathscr{E}_{\ell m}(i) \, S_{n \ell}=\mathscr{E}_{\ell m}(i) \, V_{\ell}^2 \, \alpha_{n \ell} \, .
	\end{equation}
The quantity $V_\ell^2$ is an estimate of the geometrical visibility of the total power in a multiplet $(n,\ell)$ as a function of $\ell$, whereas $\alpha_{n\ell}$ depends mainly on the frequency and excitation mechanism, i.e., $\alpha_{n\ell} \! \simeq \! \alpha(\nu_{n \ell})$. \citet[][]{DalsgaardNotes} concisely treats this issue of spatial filtering. Equation \eqref{eqn:Ballot}, however, is only strictly valid under one assumption: When the stellar flux is integrated over the full apparent disc, one must assume that the weighting function, $W$, which gives the contribution of a surface element to the integral, is a function of the distance to the disc centre alone, i.e., $W\!\!=\!\!W(\theta')$, where $\theta'$ is defined in an inertial frame with polar axis pointing toward the observer. In this case, the apparent mode amplitude can effectively be separated into two factors: $\mathscr{E}_{\ell m}(i)$ and $V_\ell^2$. This assumption holds very well in the case of intensity measurements, since the weighting function is then mainly linked to the limb-darkening, whereas for velocity measurements departures might be observed due to asymmetries in the velocity field induced by rotation \citep[see][and references therein]{BallotEtAl.2006,BallotEtAl.2008}. See Appendix~\ref{sec:Visibilities} for how to compute $\mathscr{E}_{\ell m}(i)$ and $V_\ell$.

The heights of non-radial modes are commonly defined based on the heights of radial modes according to Eq.~\eqref{eqn:Ballot}, and taking into account the $V_\ell/V_0$ ratios. Note that $\ell\!=\!0$ modes constitute a sensible reference since they are not split by rotation. Table~\ref{tab:SpatialResponses} displays the relative spatial response functions, $V_\ell/V_0$, computed according to \citet[][]{BeddingEtAl.1996}, for a number of present and upcoming instruments/missions used when measuring solar-like oscillations. Those performing intensity measurements are the red channel of the VIRGO SPM instrument on board the \emph{SOHO} spacecraft \citep[][]{VIRGO}, as well as the \emph{CoRoT} and \emph{Kepler} space missions. On the other hand, velocity measurements are performed by the HARPS spectrograph \citep[][]{HARPS} and are the purpose of the forthcoming SONG network \citep[][]{SONG}. 

	\begin{table}[H]
		\caption{Relative spatial response functions, $V_\ell/V_0$, for a number of present and upcoming instruments/missions. Notice the increased sensitivity to $\ell\!=\!3,4$ modes in velocity. Negative values of $V_\ell$ mean that the oscillations will appear to have reversed phases.}
		\centering
		\begin{threeparttable} 
		\begin{tabular}{>{$}l<{$}>{$}c<{$}>{$}c<{$}>{$}c<{$}c>{$}c<{$}>{$}c<{$}}
			\toprule
			 & \multicolumn{3}{c}{\textup{Intensity}} & \quad & \multicolumn{2}{c}{\textup{Velocity}} \\
			 & \textup{VIRGO} & \emph{CoRoT} & \emph{Kepler} && \textup{HARPS} & \textup{SONG} \\
			 & (\SI{862}{nm}) & (\SI{660}{nm}) & (\SI{641}{nm})\tnote{a} && (\SI{535}{nm}) & (\SI{550}{nm}) \\
			\midrule
			V_0/V_0 &  1.00 &  1.00 &  1.00 && 1.00 & 1.00 \\
			V_1/V_0 &  1.20 &  1.22 &  1.22 && 1.35 & 1.35 \\
			V_2/V_0 &  0.67 &  0.70 &  0.71 && 1.02 & 1.01 \\
			V_3/V_0 &  0.10 &  0.14 &  0.14 && 0.48 & 0.47 \\
			V_4/V_0 & -0.10 & -0.09 & -0.08 && 0.09 & 0.09 \\
			\bottomrule
		\end{tabular}
		\begin{tablenotes}
			\item[a] Calculated as the weighted mean over the spectral response function.
		\end{tablenotes}
		\end{threeparttable}
		\label{tab:SpatialResponses}
	\end{table}
	\FloatBarrier

Finally, a possible set of parameters going into the model is given by:
	\begin{equation}\label{eqn:Parameters}
		\vec\Theta = \tuborg{S_{n\ell},\nu_{n\ell},\Gamma_{n\ell},\rot,i,A_k,B_k,C_k,N} \, .
	\end{equation}

We have described in detail how the modelling of a power spectrum of solar-like oscillations can be achieved. When actually fitting a model to an observed power spectrum, the set of parameters entering the model might differ from the one represented in Eq.~\eqref{eqn:Parameters}. Moreover, it might be desirable to justifiably fix some of the parameters in order to reduce the dimension of parameter space.

\section{Bayesian inference}\label{sect:Bayes}
Having set up the model of the power spectrum, we will now introduce the Bayesian statistical framework to be used for estimating the model parameters and for comparing competing models.
Let us start by considering a set of competing hypotheses, $\{H_i\}$, not necessarily mutually exclusive. We should then be able to assign a probability, $p(H_i | D,I)$, to each hypothesis, taking into account the observed data, $D$, and available prior information, $I$, arising from theoretical considerations and/or previous observations. This is done through Bayes' theorem:
	\begin{equation}\label{eqn:BayesTheorem1}
		p(H_i|D,I) = \frac{p(H_i|I) p(D|H_i,I)}{p(D|I)} \, .
	\end{equation}
The probability of the hypothesis $H_i$ in the absence of $D$ is called the \emph{prior probability}, $p(H_i|I)$, whereas the probability including $D$ is called the \emph{posterior probability}, $p(H_i|D,I)$. The quantity $p(D|H_i,I)$ is called the \emph{likelihood} of $H_i$, $p(D|I)$ being the \emph{global likelihood} for the entire class of hypotheses. Bayesian inference thus encodes our current state of knowledge into a posterior probability concerning each member of the hypothesis space of interest. Moreover, the sum of the posterior probabilities over the hypothesis space of interest is unity, and thus
	\begin{equation}\label{eqn:GlobalLikelihood1}
		p(D|I)=\sum_i p(H_i|I)p(D|H_i,I) \, .
	\end{equation}

\subsection{Parameter estimation}\label{ParEst}
Very often a particular hypothesis, i.e., a model of the power spectrum, is assumed to be true and the hypothesis space of interest then relates to the values taken by the model parameters $\vec\Theta$. These parameters are continuous, which means that the quantity of interest is a PDF. The global likelihood of model $M$, assumed true, is now given by the continuous counterpart of Eq.~\eqref{eqn:GlobalLikelihood1}:
	\begin{equation}\label{eqn:GlobalLikelihood2}
		p(D|I) = \int p(\vec\Theta|I) \, p(D|\vec\Theta,I) \, \mathrm{d}\vec\Theta \, .
	\end{equation}

Let us restate Bayes' theorem in order to account for this new formalism:
	\begin{equation}\label{eqn:BayesTheorem2}
		p(\vec\Theta|D,I) = \frac{p(\vec\Theta|I) p(D|\vec\Theta,I)}{p(D|I)} \, ,
	\end{equation}
where we have substituted the hypothesis, $H_i$, with the parameters of the model that is assumed true. The terms entering this equation have the same meaning as the corresponding terms entering Eq.~\eqref{eqn:BayesTheorem1}. Use of Eq.~\eqref{eqn:BayesTheorem2} allows one to obtain the full joint posterior PDF, $p(\vec\Theta|D,I)$, this being the Bayesian solution to the problem of parameter estimation in contrast to traditional point estimation methods (e.g.~MLE). The procedure of \emph{marginalisation} makes it possible to derive the marginal posterior PDF for a subset of parameters $\vec\Theta_\mathrm{A}$, by integrating out the remaining parameters $\vec\Theta_\mathrm{B}$, called \emph{nuisance parameters}:
	\begin{equation}\label{eqn:Marginalization}
	   p(\vec\Theta_\mathrm{A}|D,I) = \int p(\vec\Theta_\mathrm{A},\vec\Theta_\mathrm{B}|D,I) \, \mathrm{d}\vec\Theta_\mathrm{B} \, .
	\end{equation}
Furthermore, assuming that the prior on $\vec\Theta_\mathrm{A}$ is independent of the prior on the remaining parameters, then by applying the product rule we have:
	\begin{equation}\label{eqn:prior}
		p(\vec\Theta_\mathrm{A},\vec\Theta_\mathrm{B}|I) = p(\vec\Theta_\mathrm{A}|I)p(\vec\Theta_\mathrm{B}|\vec\Theta_\mathrm{A},I) = p(\vec\Theta_\mathrm{A}|I)p(\vec\Theta_\mathrm{B}|I) \, .
	\end{equation} 

We will be working, in practice, with logarithmic probabilities. The global likelihood of the model plays the role of a normalisation constant and we rewrite Eq.~\eqref{eqn:BayesTheorem2} as follows:
	\begin{equation}\label{eqn:LogBayesTheorem}
		\ln p(\vec\Theta|D,I) = \mathrm{const.} + \ln p(\vec\Theta|I) + \mathscr{L}(\vec\Theta) \, .
	\end{equation}

\subsection{Model comparison}\label{ModComp}
We might also be facing a situation wherein several parametrized models are available to describe the same physical phenomenon. We then expect Bayes' theorem to allow for a statistical comparison between these competing models. In fact, Bayesian model comparison has a built-in \emph{Occam's razor}, a principle also known as \emph{lex parsimoniae}, by which a complex model is automatically penalised, unless the available data justifies its additional complexity. Notice that these might be intrinsically different models or similar models with varying number of parameters, or even the same model with different priors for its parameters.

Given two or more competing models, $\{M_i\}$, and our prior information, $I$, being in the current context that \emph{one and only one of the models is true}, we can assign individual probabilities similarly to what has been done in Eq.~\eqref{eqn:BayesTheorem1}, after substituting $H_i$ with $M_i$:
	\begin{equation}\label{eqn:BayesTheorem3}
		p(M_i|D,I)=\frac{p(M_i|I)p(D|M_i,I)}{p(D|I)} \, ,
	\end{equation}
where $p(D|M_i,I)$, also called the \emph{evidence} of model $M_i$, is given by Eq.~\eqref{eqn:GlobalLikelihood2}. The problem of model comparison is therefore analogous to the problem of parameter estimation as can be seen by comparing Eqs.~\eqref{eqn:BayesTheorem2} and \eqref{eqn:BayesTheorem3}. 

Of particular interest to us will be calculating the ratio of the probabilities of two competing models,
	\begin{equation}\label{eqn:OddsRatio}
		O_{ij} \equiv \frac{p(M_i|D,I)}{p(M_j|D,I)}=\frac{p(M_i|I)p(D|M_i,I)}{p(M_j|I)p(D|M_j,I)}=\frac{p(M_i|I)}{p(M_j|I)}B_{ij} \, ,
	\end{equation}
where $O_{ij}$ is the \emph{odds ratio} in favour of model $M_i$ over model $M_j$, $B_{ij}$ is the so-called \emph{Bayes' factor} and the remaining factor is the \emph{prior odds ratio}. We will always assume that we have no prior information impelling us to prefer one model over the other, and hence $p(M_i|I)/p(M_j|I)\!=\!1$. One is now naturally in need of a scale by which to judge the ratio of the evidences of two competing models. The usual scale employed is the Jeffreys' scale \citep[][]{Jeffreys}, which we display in Table~\ref{tab:OddsRatioScale} for convenience.   

\begin{table}[!ht]
	\centering
	\caption{Jeffreys' scale.}
	\begin{tabular}{ll}
		\toprule
		$\ln O_{ij}$ & Strength of Evidence \\
		\midrule
		<1    & Not worth more than a bare mention \\
		1--2.5   & Significant \\
		2.5--5  & Strong to very strong \\
		>5 & Decisive \\
		\bottomrule
	\end{tabular}
	\label{tab:OddsRatioScale}
\end{table}

Furthermore, the Bayesian framework makes it possible to extract parameter constraints even in the presence of model uncertainty, i.e., when the implementation of model selection has not been successful. This is done by simply combining the probability distribution of the parameters within each individual model, weighted by the model probability. This procedure, called \emph{Bayesian model averaging} \citep[see][and references therein]{Liddle}, is an analogue of the superposition of eigenstates of an observable in quantum mechanics.

\subsection{Ignorance priors}
The main advantage of the Bayesian framework when compared to a frequentist approach is the ability to incorporate relevant prior information through Bayes' theorem and evaluate its effect on our conclusions. Assuming that the prior on each parameter is independent of the prior on any other parameter, then according to Eq.~\eqref{eqn:prior} we have:
	\begin{equation}\label{eqn:logprior}
		p(\vec\Theta|I) = \prod_k f_k(\Theta_k) \, ,
	\end{equation}
where $f_k(\Theta_k)$ is the prior PDF associated with the $k$th parameter entering the model. As our state of knowledge of a particular physical phenomenon evolves through continued study and experimentation, the set of priors relevant for the analysis of a new data set will change. In the early stages of research, however, we look for a set of priors that encode our rather limited state of knowledge, i.e., a set of \emph{ignorance priors} \citep[e.g.,][and references therein]{Gregory2005}.

When dealing with \emph{location parameters}, e.g.~$\{\nu_{n\ell}\}$ in Eq.~\eqref{eqn:Parameters}, our choice of prior would at first be the \emph{uniform prior}:
	\begin{align}
		f_k(\Theta_k) &= \begin{cases}
			\frac{1}{\Theta_k^\textup{max} - \Theta_k^\textup{min}} & \text{, for } \Theta_k^\textup{min} \leq \Theta_k \leq \Theta_k^\textup{max} \, , \\
			0 &\text{, otherwise} \, .
		\end{cases}
	\end{align}
If we are ignorant about the limits $\Theta_k^\textup{min}$ and $\Theta_k^\textup{max}$, then we refer to $f_k(\Theta_k)$ as an improper prior, meaning that it is not normalised. An improper prior is not suitable for model comparison problems. On the other hand, when dealing with \emph{scale parameters}, e.g.~$\{S_{n\ell}\}$ in Eq.~\eqref{eqn:Parameters}, our choice of prior might be that of a \emph{Jeffreys' prior}:
	\begin{align}
		f_k(\Theta_k) &= \begin{cases}
			\frac{1}{\Theta_k \ln\kant[\big]{\Theta_k^\textup{max} / \Theta_k^\textup{min}}} & \text{, for } \Theta_k^\textup{min} \leq \Theta_k \leq \Theta_k^\textup{max} \, , \\
			0 &\text{, otherwise} \, .
		\end{cases}
	\end{align}
By employing a Jeffreys' prior we are assigning equal probability per decade (scale invariance), mainly useful when the prior range spans several orders of magnitude. In case the prior lower limit includes zero, a \emph{modified Jeffreys' prior} should be used instead to avoid the divergence at zero:
	\begin{align}\label{eqn:ModJeffreys}
		f_k(\Theta_k) &= \begin{cases}
				\frac{1}{ \p[\big]{\Theta_k+\Theta_k^\textup{uni}} \ln\kant[\big]{\p[\big]{\Theta_k^\textup{uni}+\Theta_k^\textup{max}} / \Theta_k^\textup{uni}}} &\text{, for } 0 \leq \Theta_k \leq \Theta_k^\textup{max} \, , \\
			0 &\text{, otherwise} \, .
		\end{cases}
	\end{align}
For $\Theta_k\!\!\gg\!\!\Theta_k^\textup{uni}$, Eq.~\eqref{eqn:ModJeffreys} behaves just like a Jeffreys' prior, whereas for $\Theta_k\!\ll\!\Theta_k^\textup{uni}$ it behaves like a uniform prior, thus not diverging at zero. $\Theta_k^\textup{uni}$ marks the transition between the two regimes.

\section{Markov chain Monte Carlo}\label{sect:MCMC}
After inspection of Eq.~\eqref{eqn:Marginalization}, the need for a mathematical tool that is able to efficiently evaluate the multi-dimensional integrals required in the computation of the marginal posteriors becomes clear. This constitutes the rationale behind the method known as Markov chain Monte Carlo, first introduced in the early 1950s by statistical physicists and nowadays widely used in all areas of science and economics.

\subsection{Metropolis--Hastings algorithm}\label{sect:M-H}
The aim is to draw samples from the \emph{target distribution}, $p(\vec\Theta|D,I)$, by constructing a pseudo-random walk in model parameter space such that the number of samples drawn from a particular region is proportional to its posterior density. Such a pseudo-random walk is achieved by generating a Markov chain, whereby a new sample, $X_{t+1}$, depends on the previous sample{\footnote{A remark on the notation: $X_{t}$ may be thought of as a single vector in parameter space.}}, $X_{t}$, in accordance with a time-independent quantity called the \emph{transition kernel}, $p(X_{t+1}|X_{t})$. After a burn-in phase, $p(X_{t+1}|X_{t})$ is able to generate samples of $\vec\Theta$ with a probability density converging on the target distribution. The Markov chain must fulfil three requirements in order to achieve this convergence: it must be \emph{irreducible}, \emph{aperiodic} and \emph{positive recurrent} \citep[][]{Roberts96}.

The algorithm that we employ in order to generate a Markov chain was initially proposed by \citet[][]{Metropolis}, and subsequently generalised by \citet[][]{Hastings}, this latter version being commonly referred to as the Metropolis--Hastings algorithm.
It works in the following way: Suppose the current sample, at some instant denoted by $t$, is represented by $X_t$. We would like to steer the Markov chain toward the next sampling state, $X_{t+1}$, by first proposing a new sample to be drawn, $Y$, from a \emph{proposal distribution}, $q(Y|X_t)$, centred on $X_t$. Here we specifically treat $q(Y|X_t)$ as being a multivariate normal distribution with covariance matrix $\tens{\Sigma}$. We employ independent Gaussian parameter proposal distributions and thus $\tens{\Sigma}$ is assumed diagonal. The proposed sample is then accepted with a probability given by:
	\begin{equation}
		\alpha(X_t,Y) = \min(1,r) = \min\kant*{1,\frac{p(Y|D,I)}{p(X_t|D,I)} \frac{q(X_t|Y)}{q(Y|X_t)}} \, ,
	\end{equation}
where $\alpha(X_t,Y)$ is the \emph{acceptance probability} and $r$ is called the \emph{Metropolis ratio}. In the present case of a symmetric proposal distribution, we have $q(X_t|Y)\!=\!q(Y|X_t)$. As a result, if the posterior density for the proposed sample is greater than or equal to that of the current sample, i.e., $p(Y|D,I) \! \geq \! p(X_t|D,I)$, then the proposal will always be accepted, otherwise it will be accepted with a probability given by the ratio of the posterior densities. If $Y$ is not accepted, then the chain will keep the current sampling state, i.e., $X_{t+1}\!\!=\!\!X_t$. The procedure just described is repeated for a predefined number of iterations or, alternatively, for a number of iterations determined by a convergence test applied to the Markov chain \citep[e.g.,][]{GelmanRubin}. The total number of iterations is denoted by $n_\textup{it}$.

Once a Markov chain has been created, the problem of marginalization becomes trivial, as the way to extract information on the individual parameters is simply to generate a histogram for each parameter and thus obtain its PDF.
An appropriate number of bins in the histograms can be selected using for example Scott's criterion \citep{ScottCriterion}.
Usually the information in the PDF will be condensed using some summary statistics, like for example finding the median of the distribution and the 68\% credible region around it.

	\begin{figure}[tb]
		\centering
		\includegraphics[width=0.45\textwidth]{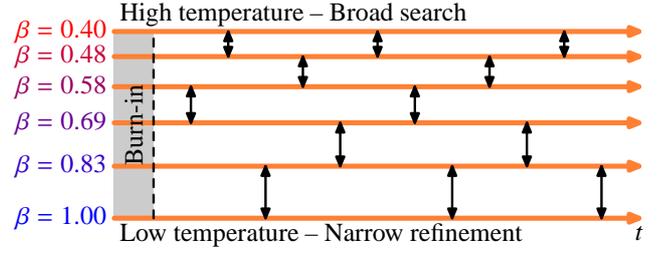}
		\caption{Schematic representation of the functioning of the parallel tempering mechanism, whereby tempering chains are allowed to swap their parameter states (swaps are indicated by vertical arrows).}
		\label{fig:ParallelTempering}
	\end{figure}

\subsection{Parallel tempering}\label{sec:ParallelTempering}
The Metropolis--Hastings algorithm outlined above might become stuck in a local maximum of the target distribution, thus failing to fully explore all regions in parameter space containing significant probability. A way of overcoming this is to employ parallel tempering \citep[e.g.,][]{EarlDeem}, whereby a discrete set of progressively flatter versions of the target distribution are created by introducing a \emph{temperature parameter}, $\mathscr{T}$.
In practice, use is made of its reciprocal, $\beta\!=\!1/\mathscr{T}$, referred to as the \emph{tempering parameter}.
By modifying Eq.~\eqref{eqn:BayesTheorem2}, we generate the tempered distributions as follows:
	\begin{equation}
		p(\vec\Theta|D,\beta,I) \propto p(\vec\Theta|I) \, p(D|\vec\Theta,I)^\beta \, , \quad 0 < \beta \leq 1 \, .
	\end{equation}
For $\beta\!=\!1$, we recover the target distribution, also called the \emph{cold sampler}, whereas for $\beta\!<\!1$ the hotter distributions are effectively flatter versions of the target distribution. Drawing samples from a hotter, i.e., flatter, version of the target distribution will allow, in principle, to visit regions of parameter space containing significant probability, otherwise not accessible to the basic algorithm. The problem of parameter estimation obviously continues to rely on samples drawn from the cold sampler. In Sect.~\ref{sect:ModelComparison} we describe how samples drawn from the remaining tempered distributions are useful in evaluating Bayes' factor.

Implementation of parallel tempering works in the following way: Several versions of the Metropolis--Hastings algorithm are launched in parallel ($n_\beta$ in total), each being characterised by a different tempering parameter, $\beta_i$. At random intervals, comprehending a mean number ($n_\textup{swap}$) of iterations, a pair of adjacent chains, labelled with $\beta_i$ and $\beta_{i+1}$, is randomly chosen and a proposal is made to swap their parameter states. The proposed swap is then accepted with a probability given by:
	\begin{equation}
	\begin{split}
		 \alpha_\textup{swap} &= \min(1,r_\textup{swap}) \\
		 &= \min\kant*{1, \frac{p(X_{t,i+1}|D,\beta_i,I) p(X_{t,i}|D,\beta_{i+1},I)}{p(X_{t,i}|D,\beta_i,I) p(X_{t,i+1}|D,\beta_{i+1},I)}} \, ,
	\end{split}
	\end{equation}
where, at instant $t$, chain $\beta_i$ is in state $X_{t,i}$ and chain $\beta_{i+1}$ is in state $X_{t,i+1}$. By running such a set of cooperative chains, we effectively enable the algorithm to sample the target distribution in a way that allows for both the investigation of its overall features (low-$\beta$ chains) and the examination of the fine details of a local maximum (high-$\beta$ chains). A schematic representation of the functioning of the parallel tempering mechanism is shown in Fig.~\ref{fig:ParallelTempering}. In Fig.~\ref{fig:PseudoCode}, a version of the Metropolis--Hastings algorithm is shown, written in pseudocode, and with the inclusion of the parallel tempering mechanism.

Concerning the values taken by the tempering parameter, $\{\beta_i\}$, optimal values are chosen in order to achieve a \emph{swap acceptance rate} between adjacent levels of $\sim\!50\%$. Heuristically, we can assert that by employing a geometric progression \citep[cf.][]{Benomar09},
	\begin{equation}
		\beta_i = \lambda^{1-i} \, ,
	\end{equation}
such a desideratum is reached by setting $\lambda \! \sim \! 1.2$. 
The number of chains, $n_\beta$, should be chosen such as to reach a desired balance between sampling efficiency and computational time. However, when using the parallel tempering mechanism in model comparison problems, as we will get back to in Sect.~\ref{sect:ModelComparison}, a large number of tempering chains are needed (typically $n_\beta \gtrsim 10$). The value of $n_\textup{swap}$ should be chosen inversely proportional to $n_\beta$ (typically a few dozens).

\subsection{Automated MCMC}

\begin{figure}[tb]
	\begin{framed}
	\begin{algorithmic}[1]
		\Procedure{Parallel Tempering Metropolis--Hastings}{}
			\State $X_{0,i} = X_0 \, , \; 1 \leq i \leq n_\beta$
			\For{$t = 0,1,\ldots,n_\textup{it}-1$}
				\For{$i = 1,2,\ldots,n_\beta$}
					\State Propose a new sample to be drawn from a
					\Statex \hspace{5em} proposal distribution: $Y \sim N(X_{t,i};\tens{\Sigma}_i)$
					\State Compute the Metropolis ratio:
					\Statex \hspace{5em} $\ln r = \ln p(Y|D,\beta_i,I) - \ln p(X_{t,i}|D,\beta_i,I)$
					\State Sample a uniform random variable:
					\Statex \hspace{5em} $U_1\sim\textup{Uniform}(0,1)$
					\If{$\ln U_1 \leq \ln r$}
						\State $X_{t+1,i} = Y$
					\Else
						\State $X_{t+1,i} = X_{t,i}$
					\EndIf
				\EndFor
				\State $U_2 \sim\textup{Uniform}(0,1)$
				\If{$U_2 \leq 1/n_\textup{swap}$}
					\State Select random chain: 
					\Statex \hspace{5em} $i\sim \textup{UniformInt}(1,n_\beta-1)$ 
					\State Compute $r_\textup{swap}$:
					\Statex \hspace{5em} $\ln r_\textup{swap} = \ln p(X_{t,i+1}|D,\beta_i,I) + \ln p(X_{t,i}|D,\beta_{i+1},I)$
					\Statex \hspace{6em} $- \ln p(X_{t,i}|D,\beta_i,I) - \ln p(X_{t,i+1}|D,\beta_{i+1},I)$
					\State $U_3 \sim\textup{Uniform}(0,1)$
					\If{$\ln U_3 \leq \ln r_\textup{swap}$}
						\State Swap parameter states of chains $i$ and $i+1$:
						\Statex \hspace{7em} $X_{t,i} \leftrightarrow X_{t,i+1}$
					\EndIf
				\EndIf
			\EndFor
			\State \Return $X_{t,i} \, , \; \forall t \, , \; i\!:\!\beta_i\!=\!1$
		\EndProcedure
	\end{algorithmic}
	\end{framed}
	\caption{Version of the Metropolis--Hastings algorithm written in pseudocode and with the inclusion of parallel tempering.}
	\label{fig:PseudoCode}
\end{figure}

\begin{figure}
		\centering
		\subfloat{\includegraphics[width=0.24\textwidth]{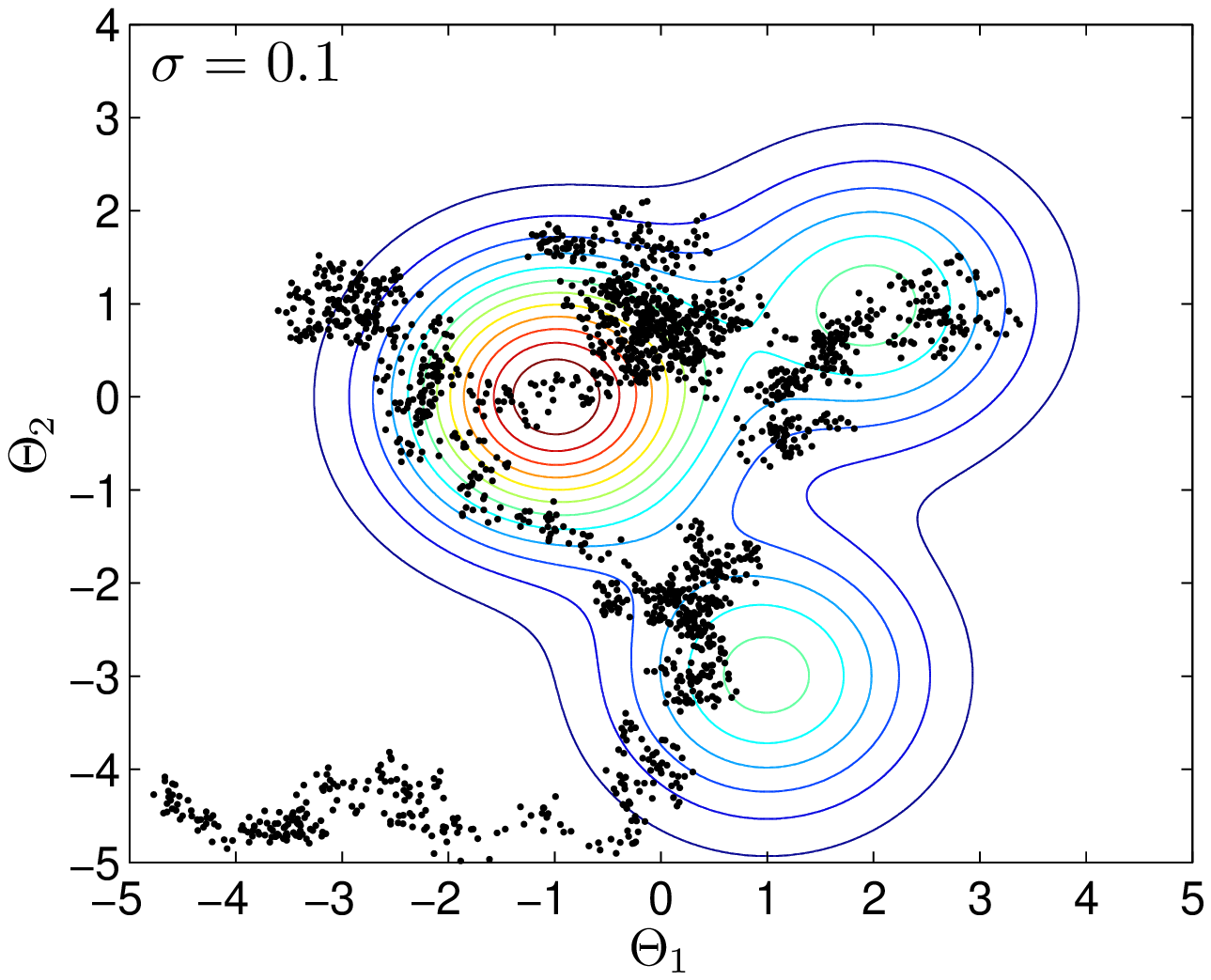}}
		\subfloat{\includegraphics[width=0.24\textwidth]{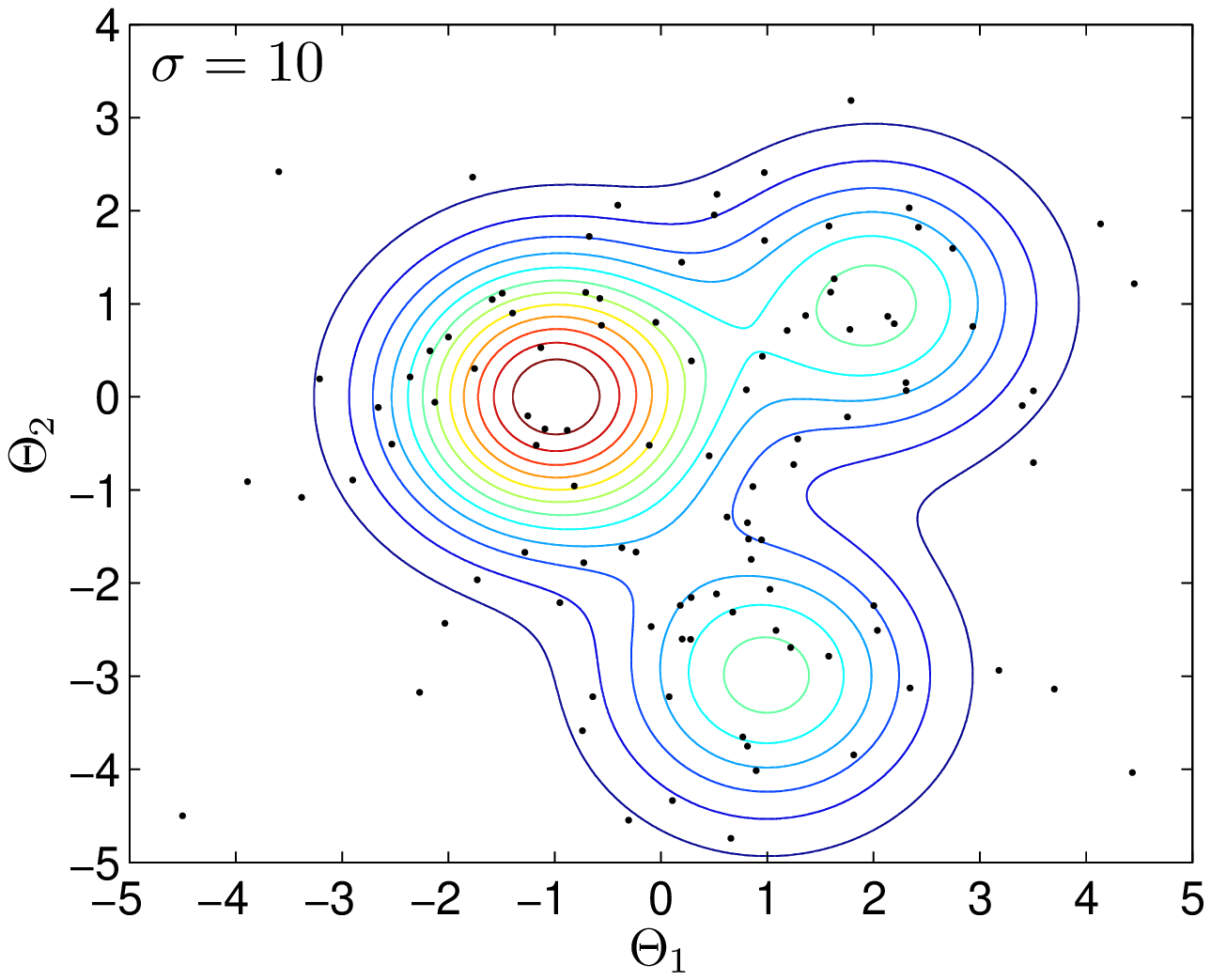}} \\
		\includegraphics[width=0.24\textwidth]{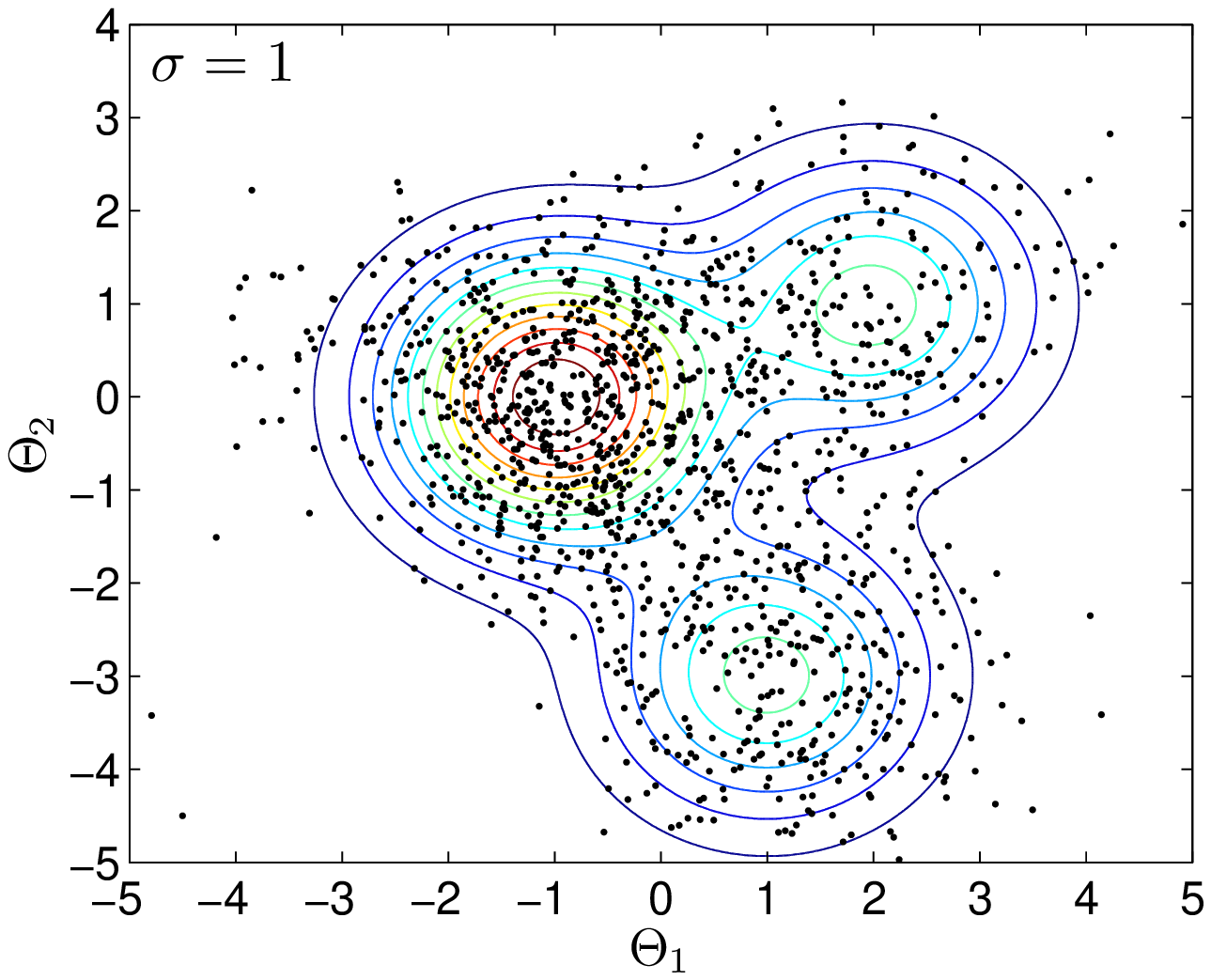}
		\caption{The same target distribution sampled by three chains, each being characterised by a different set $\{\sigma\}$. The contours map the target distribution, which in turn depends only on the parameters $\Theta_1$ and $\Theta_2$. The starting point in each of the chains is $(\Theta_1,\Theta_2)\!=\!(-4.5,-4.5)$ and all contain $2000$ iterations. Both parameters share a common $\sigma$-value whose optimal setting is $\sigma\!=\!1$. It is important to note that given a sufficiently large number of iterations, all the chains would eventually map out the target distribution, however an optimal choice of the proposal distribution will result in significantly faster convergence.}
		\label{fig:AcceptRate}
	\end{figure}

So far we have not mentioned the need to adequately choose the set $\{\sigma\}$ of diagonal elements of the $\mat{\Sigma}$ matrix, indicating the width of the Gaussian proposal distribution for each parameter. The set of individual $\sigma$ values specifies the direction and step size in parameter space when proposing a new sample to be drawn. The optimal choice of $\{\sigma\}$ is closely related to the average rate at which proposed state changes are accepted, the so-called \emph{acceptance rate}. Accordingly, small $\sigma$ values will lead to a large acceptance rate, with successive samples being highly correlated and ultimately requiring a large number of iterations in order to yield equilibrium distributions of model parameters. On the other hand, large $\sigma$ values will lead to a low acceptance rate, meaning that proposed state changes will seldom be accepted.
This is illustrated in Fig.~\ref{fig:AcceptRate}, where the same simplified target distribution is sampled by three chains, each being characterised by a set of $\sigma$ values differing on the respective magnitudes. \citet[][]{RobertsEtAl.1997} recommend, based on empirical studies, calibrating the acceptance rate to $\sim\!25\%$ when dealing with a high-dimensional model as it is the case when performing a global peak-bagging.

One could, of course, employ a trial-and-error approach and manually calibrate the $\sigma$ values. However, since we are dealing with a large number of parameters that, in addition, correspond to several different physical quantities, this would quickly become very time-consuming and impractical. We instead employ an automated process of calibration of the proposal $\sigma$ values, which is based on a statistical control system similar to the one described in \citet[][]{GregoryBook}. The control system makes use of an error signal to steer the selection of the $\sigma$ values during the burn-in stage of a single parallel tempering MCMC run, acting independently on each of the tempered chains. The error signal is proportional to the difference between the current acceptance rate and the target acceptance rate. As soon as the error signal for each of the tempered chains is less than a measure of the Poisson fluctuation expected for a zero mean error(computed as the square root of the target acceptance rate times the number of iterations between changes in the $\sigma$ values), the control system is turned off and the algorithm switches to the standard parallel tempering MCMC. In practice this effectively marks the end of the burn-in stage.

The control system as briefly described here is also used in \citet[][]{Gruberbauer}, whereas \citet[][]{Benomar09} employ a self-learning process that appropriately adapts the covariance matrix, assumed non-diagonal.   

\subsection{Model comparison using parallel tempering MCMC}\label{sect:ModelComparison}
We are now interested in computing the odds ratio, $O_{ij}$, in favour of model $M_i$ over model $M_j$ according to Eq.~\eqref{eqn:OddsRatio}. When analysing solar-like oscillations, a recurrent difficulty is to correctly tag the modes of oscillation by angular degree $\ell$. There are two possible ways of tagging the modes or, equivalently, two competing models. Computation of $O_{ij}$ is thus a means of assessing which of the two identification scenarios is statistically more likely (although not necessarily physically more meaningful, as is often misinterpreted).  

Samples drawn from the tempered distributions can, in principle, be used to compute the global likelihood, $p(D|M_i,I)$, of a given model $M_i$. Notice that Bayes' factor, $B_{ij}$, is defined as the ratio of the global likelihoods of two competing models:
	\begin{equation}
		B_{ij} \equiv \frac{p(D|M_i,I)}{p(D|M_j,I)} = \exp\kant*{ \ln p(D|M_i,I) - \ln p(D|M_j,I) } \, .
	\end{equation}
It can be shown that the global likelihood of a model is given by \citep[for a derivation see][]{GregoryBook}:
	\begin{equation}\label{eqn:GlobalLikelihood}
		\ln p(D|M_i,I) = \int_0^1 \inner{ \ln p(D|M_i,X,I) }_\beta \, \mathrm{d}\beta \, ,
	\end{equation}
where
	\begin{equation}
		\inner{ \ln p(D|M_i,X,I) }_\beta = \frac{1}{n} \sum_t \ln p(D|M_i, X_{t,\beta}, I) 
	\end{equation}
is the expectation value of the natural logarithm of the likelihood for a particular tempered chain characterised by $\beta$. The set $\{X_{t,\beta}\}$ represents the corresponding samples drawn after the burn-in stage, while $n$ is the number of samples in each set. A sufficient number ($\gtrsim$10) of parallel tempered chains is required if we are to estimate the integral in Eq.~\eqref{eqn:GlobalLikelihood} by interpolating values of $\inner{ \ln p(D|M_i,X,I) }_\beta$.

\section{Examples}\label{sect:Application}
In the following we will pick a couple of examples where we have applied the described Automated Parallel Tempering MCMC formalism to recent measurements of solar-like oscillators.

\subsection{HD~49933: The importance of priors}\label{sec:HD49933}
	\begin{figure*}[t]
		\centering
		\includegraphics[width=0.97\textwidth]{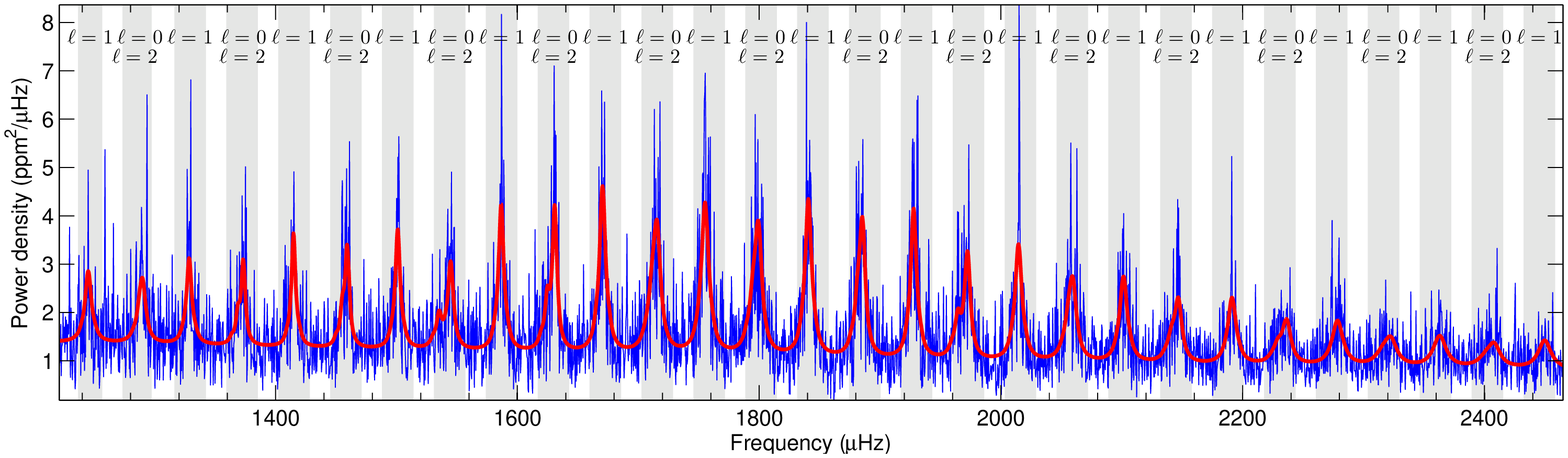}
		\caption{Power density spectrum of HD~49933 with the best-fitting model (using prior set S2) overlaid. The shaded areas indicates the ranges of the uniform priors on the frequencies.}
		\label{fig:FinalSolution}
	\end{figure*}
We have performed an analysis of the star HD~49933, based on 180 days of photometry from the \emph{CoRoT} satellite arising from two runs: The initial 60-day run, IRa01, and 120 days from the longer second run, LRa01. The time series was split up into segments of 30 days and the power spectra of the individual segments were averaged to construct a mean power spectrum ($s\!=\!6$ in Eq.~\ref{eqn:StatisticsAveraged}). The acoustic spectrum of this F5 main-sequence star has proven to be very difficult to interpret mainly due to the relatively large linewidths (see Fig.~\ref{fig:FinalSolution}). We assume the ridge identification denoted as ``Scenario B'' in \cite{BenBauCam}.

The acoustic spectrum was fitted using the APT~MCMC formalism, but using two different sets of priors (see Table~\ref{tab:HD49933Priors}).
The first set (S1) was constructed using only ignorance priors, while the second set (S2) includes knowledge about the stellar rotation.
From spectroscopic and asteroseismic studies of HD~49933, \cite{Bruntt} was able to constrain the rotation of the star to $v\sin(i) = \SI{10 \pm 1}{km~s^{-1}}$ and the radius to $R/\Rsun = 1.385 \pm 0.031$, which can be combined to impose a constraint on the projected rotational splitting, $\rot^* = \rot\sin(i)$, of $1.65\pm0.17$ $\mu$Hz. In set S2 this knowledge is added as a gaussian prior on the projected splitting of the star.
In both cases the fits were done using the following configuration:
\begin{itemize}
\item 15 orders were fitted with $\ell=0,1,2$ modes in a fitting window spanning from 1220 to $\SI{2465}{\mu Hz}$ (see Fig.~\ref{fig:FinalSolution}).
\item One linewidth and one height per order assigned to the $\ell\!=\!0$ mode, and then linearly interpolated by frequency and scaled to the higher degree modes.
\item Rotation and inclination angle fitted with the two free parameters, $\rot^*$ and $i$.
\item The background was parametrized as a sum of 3 Harvey-like models plus a white noise contribution.
\item \num{800000} samples were drawn from the target distribution, employing 10 parallel chains.
\end{itemize}

\begin{figure}[!ht]
		\centering
		\subfloat[\label{fig:PriorsAreImportant}]{
			\begin{minipage}[c]{0.24\textwidth}
				\includegraphics[width=\textwidth]{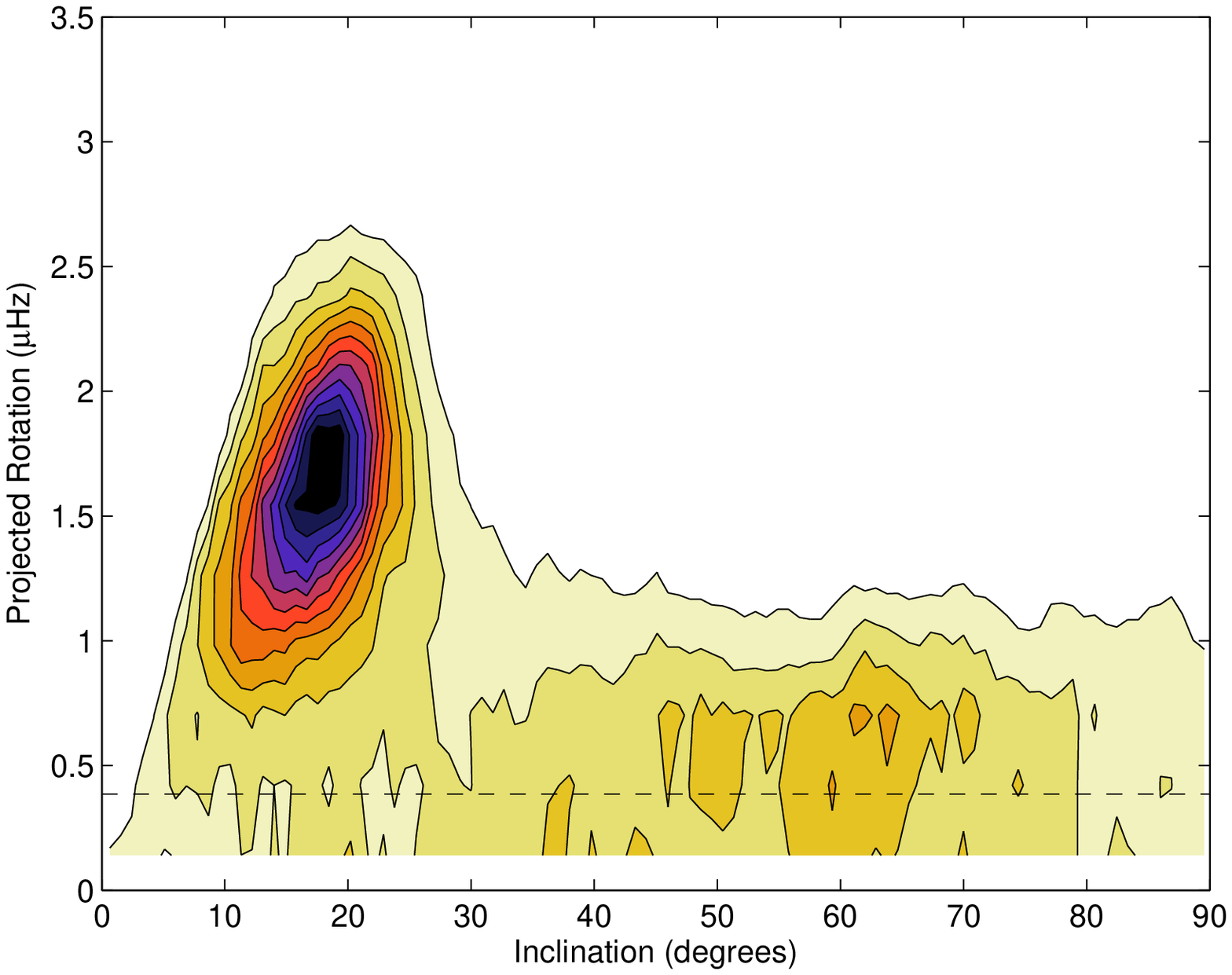} \\
				\includegraphics[width=\textwidth]{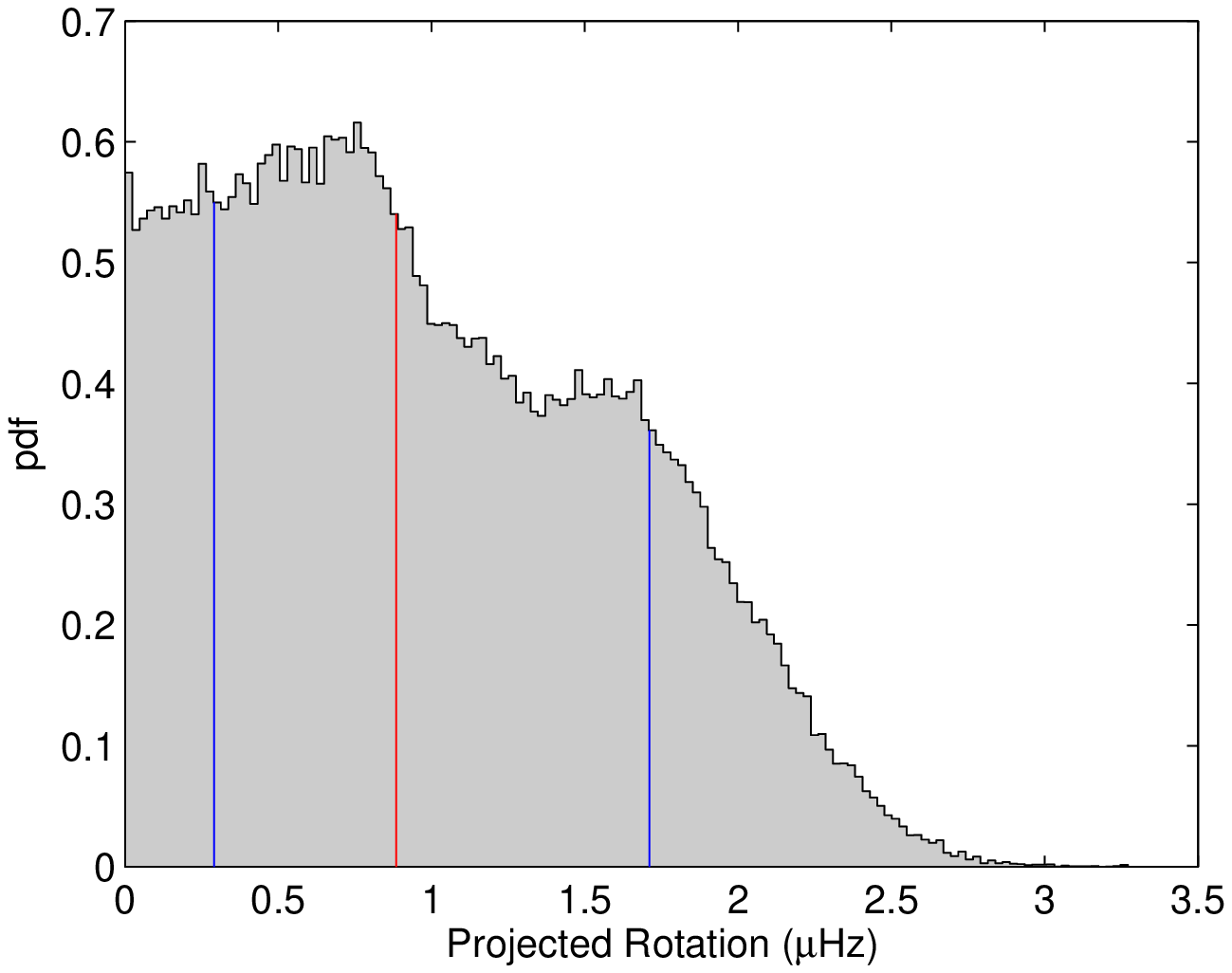} \\
				\includegraphics[width=\textwidth]{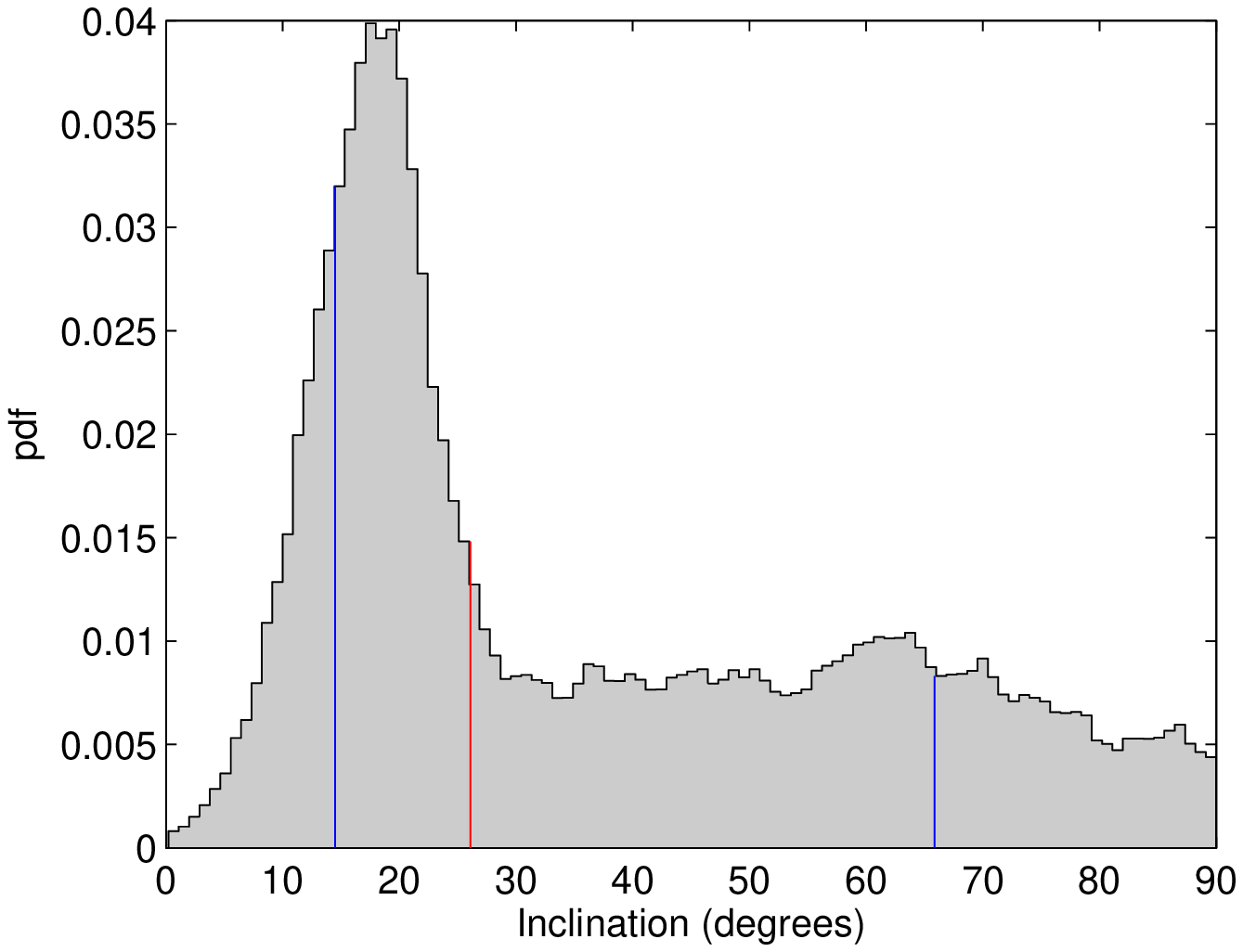}
			\end{minipage}
		}
		\subfloat[\label{fig:PriorsAreImportant2}]{
			\begin{minipage}[c]{0.24\textwidth}
				\includegraphics[width=\textwidth]{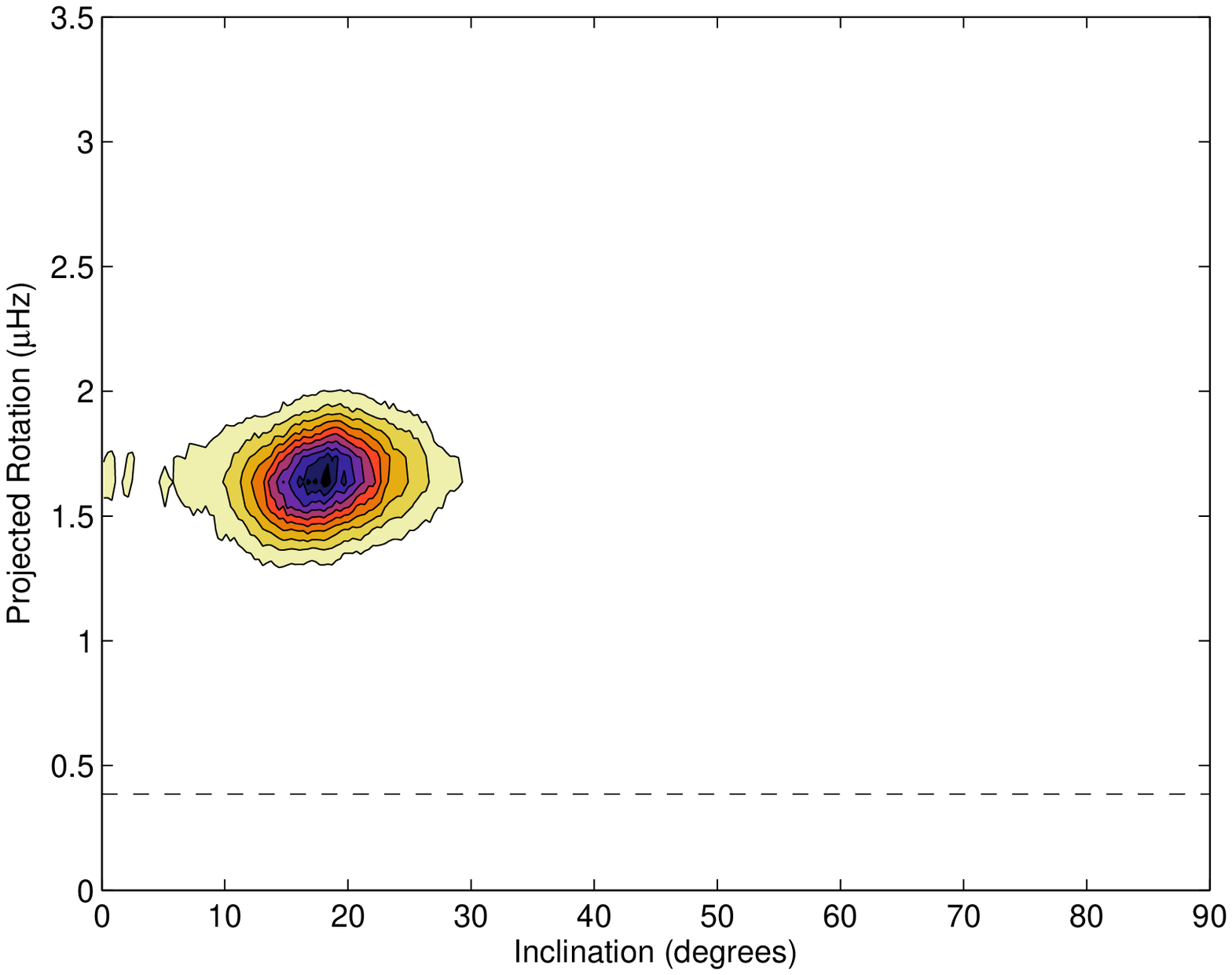} \\
				\includegraphics[width=\textwidth]{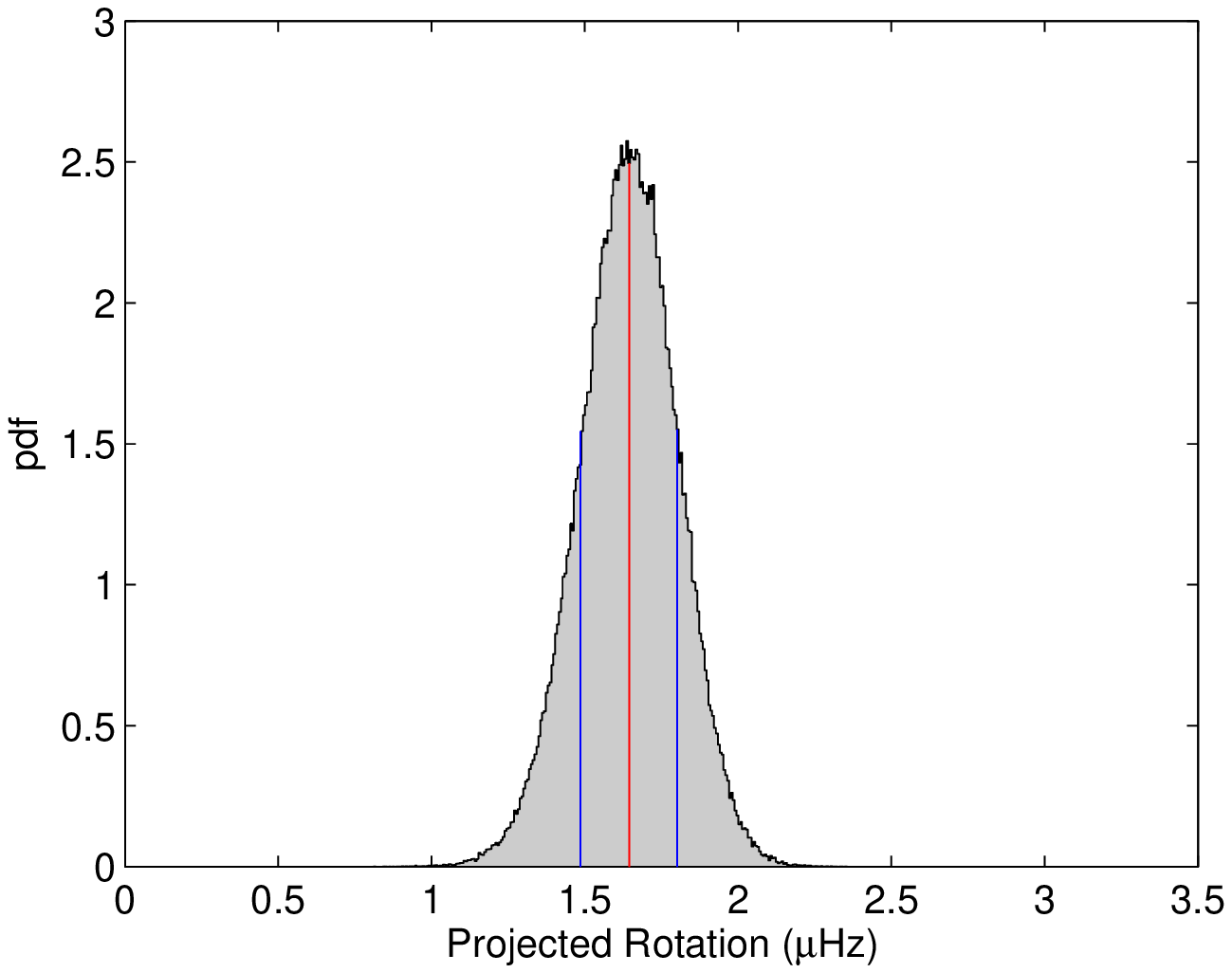} \\
				\includegraphics[width=\textwidth]{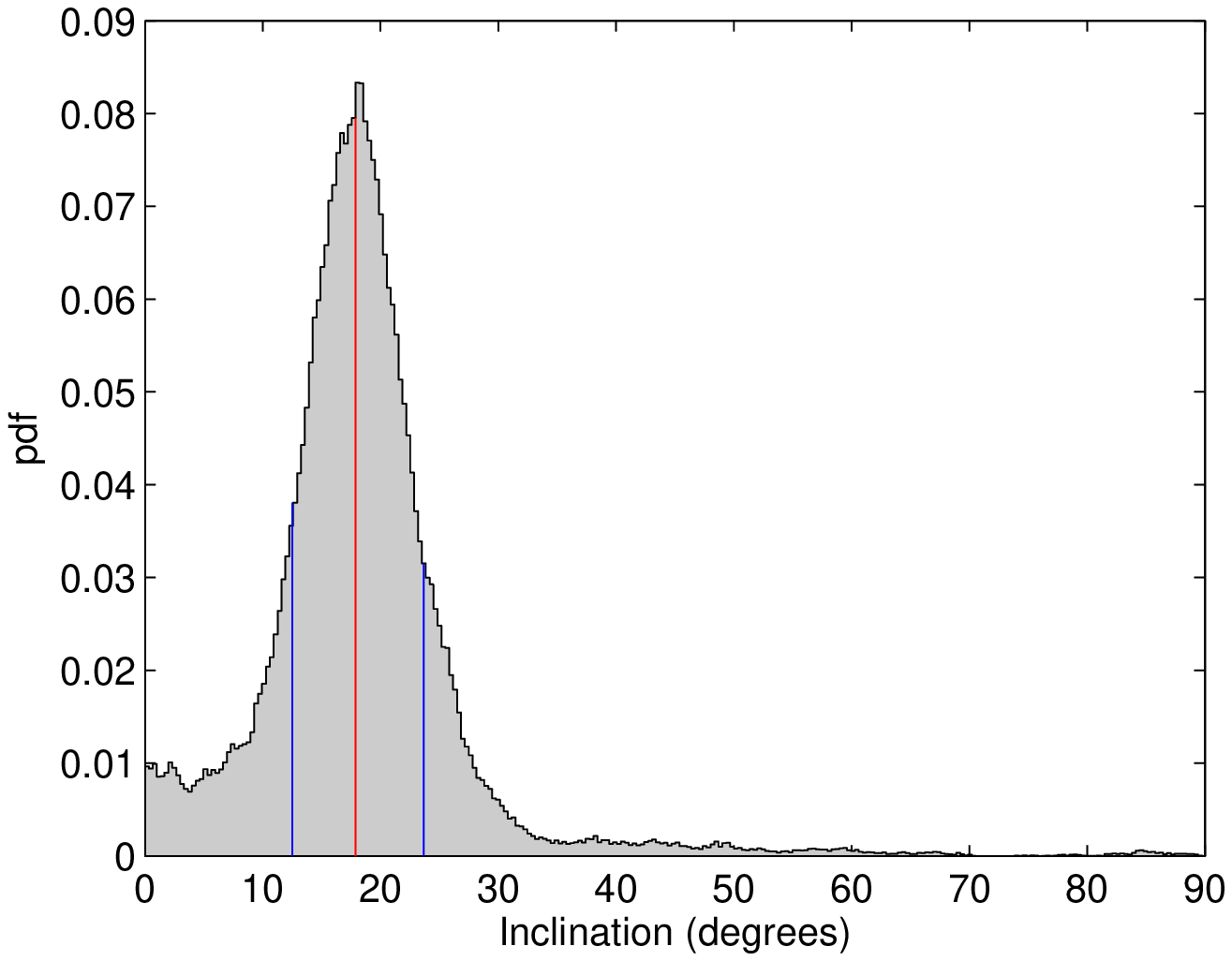}
			\end{minipage}
		}
		\caption{Results on the rotation and inclination angle of HD~49933. In both cases the prior on the inclination angle is uniform in the interval $0\degree$--$90\degree$. \protect\subref{fig:PriorsAreImportant} We employ a uniform prior on rotation (S1). \protect\subref{fig:PriorsAreImportant2} We employ a Gaussian prior on projected rotation (S2). The vertical lines in the histograms indicate the median and the boundaries of the 68\% credible regions of the distributions. The dashed line in the top figures indicates the frequency resolution in the spectrum.}
	\end{figure}

First of all, it is important to note that the results are consistent with the ones reported in \cite{BenBauCam}. For example the derived frequencies and linewidths are all well within the error bars. We will here focus on the results of the rotational splitting and inclination angle.
The probability density functions for the fitted parameters when using ignorance priors (S1) are shown in Fig.~\ref{fig:PriorsAreImportant} and, after applying the Gaussian prior on the projected splitting (S2), the results change to the ones shown in Fig.~\ref{fig:PriorsAreImportant2}.

	\begin{table}
		\caption{Prior input for the HD~49933 analysis.}
		\centering
		\begin{tabular}{ll}
			\toprule
			Parameter & Prior \\
			\midrule
			Frequencies & Uniform \\
			Heights & Modified Jeffreys \\
			Linewidths & Uniform \\
			Inclination & Uniform ($0\degree$--$90\degree$) \\
			Rotation & S1: Uniform on $\rot$ (0--$\SI{10}{\mu Hz}$) \\
			         & S2: Gaussian on $\rot^*$ ($1.65\pm\SI{0.17}{\mu Hz}$) \\
			\bottomrule
		\end{tabular}
		\label{tab:HD49933Priors}
	\end{table}

What these results demonstrate is, first of all, that these techniques are extremely efficient at probing and constraining parameters which traditional methods would have considerable difficulty in constraining.
In \cite{BenBauCam}, where similar techniques are employed, the derived inclination angle was ${17\degree} {}^{+7\degree}_{-9\degree}$ and the rotational splitting placed in the range 3.5--\SI{6.0}{\mu Hz}, with which our results are perfectly consistent with.
Another point to be drawn from this example is the importance of the inclusion of prior knowledge. By incorporating our prior knowledge about the rotation of the star through the accurate measurements from a spectral analysis, we are able to yield a much cleaner constraint on the inclination angle, which in Fig.~\ref{fig:PriorsAreImportant} has a considerable tail of probability towards large inclinations.

It is important to note that this prior on $\rot^*$ is strong in the sense that it dominates the fit. This simply comes from the fact that the data do not provide any further information on this parameter and so our prior knowledge of the model is still providing the best constraint.
In such cases one of course has to be careful that such a strong prior is not wrongfully restrictive.
If other effects (in this case, for instance, differential rotation or second-order rotational effects) were present, this could introduce biases to the fitted results.
This could of course be tested using the methodology described in Sects.~\ref{ModComp} and \ref{sect:ModelComparison}, by constructing models that incorporate these effects and testing their significance.

\subsection{Procyon: The problem of ridge identification}\label{sec:Procyon}
Here we address the issue of tagging the oscillation modes by angular degree in the case of the F5 star Procyon. We have thus reanalysed the data acquired during a multi-site campaign \citep{ProcyonI,ProcyonII} carried to observe oscillations in this star. The data consist of high-precision velocity observations obtained over more than three weeks with eleven telescopes, representing the most extensive campaign organised so far on a solar-type oscillator.

The problem of ridge identification in F stars dates back to when \emph{CoRoT} observations of HD~49933 were first analysed \citep{Appourchaux08}, a problem that would be recently solved for this star only after a new longer time series was made available \citep{BenBauCam}. \citet{ProcyonII} address this same problematic in the case of Procyon by employing three distinct methodologies: (i) a collapsed power spectrum along several radial orders, (ii) a scaled \'echelle diagram \citep{ScaledEchelle} and (iii) Bayesian model comparison (as described in Sect.~\ref{sect:ModelComparison}). The last-mentioned methodology statistically favours their Scenario A over their Scenario B identification, whereas the first and second methodologies suggest the contrary although without quantifying their preference for Scenario B in a statistical sense. 

We performed a peak-bagging of the power spectrum of Procyon considering both identification scenarios while simultaneously testing for the presence of $\ell\!=\!3$ modes. This gives a total of four competing models, i.e., $\{M_{\rm{A}},M_{\rm{A}}^{\ell=3},M_{\rm{B}},M_{\rm{B}}^{\ell=3}\}$, the notation chosen to be unambiguous. The details of the peak-bagging as implemented here slightly differ from those presented in \citet{ProcyonII}, and especially concern the limits of the fitting window and the way in which the background was parametrized. The details are as follows:
\begin{itemize}

\item The peak-bagging was performed on the sidelobe-optimised power density spectrum whose intrinsic frequency resolution is $0.77\:{\rm{\mu Hz}}$. Peaks were described by symmetric Lorentzians centred on the mode frequencies. Three frequencies were fitted per overtone, each with a different angular degree ($\ell\!=\!0,1,2$).
Type of prior imposed at first: independent and uniform, centred ($\pm\SI{8}{\mu Hz}$) on the initial guesses. The mode frequencies were further constrained to lie close to the ridge centroids and to have only small jumps from one order to the next. Also, a Gaussian prior ($\mu\!=\!\SI{4}{\mu Hz}$, $\sigma\!=\!\SI{5}{\mu Hz}$) was imposed on the small frequency separation, $\delta\nu_{02}$, between adjacent modes with $\ell\!=\!0$ and $\ell\!=\!2$. The small separation was not itself a free parameter in the fit, but instead a derived quantity. Note that the last-mentioned constraint implies that the type of prior on the $\ell\!=\!0,2$ frequency parameters is ultimately not independent nor uniform.
Optionally, modes with $\ell\!=\!3$ could be included in the model with their frequencies fixed to
	\begin{equation}
		\nu_{n-1,3} = \nu_{n,1} - \tfrac{5}{3}(\nu_{n,0}-\nu_{n-1,2}) \, ,
	\end{equation}
according to the asymptotic relation \citep{Tassoul}.
A total of 14 overtones were considered and the fitting window runs from 500 to $\SI{1300}{\mu Hz}$. By employing this construction it is assumed that no mixed modes are present in the fitting window.
The inclusion of $\ell\!=\!3$ modes does not add any more free parameters, while adding however their features to the model spectrum which are derived from the fitted $\ell=0,1,2$ frequencies.

\item The linewidth was parametrized as a linear function of frequency, defined by two parameters $\Gamma_{600}$ and $\Gamma_{1200}$, which are the values at 600 and $1200\:{\rm{\mu Hz}}$, respectively. Both parameters were fitted. Type of prior imposed: uniform in the range 0--$10\:{\rm{\mu Hz}}$.

\item The heights of radial modes in units of power density were fixed according to \citet{ChaplinEtAl2008}:
\begin{equation}
	S_{n0}=\frac{2 \, A^2 \, T}{\pi T \Gamma_{n0} +2} \, ,	
\end{equation}
where $A^2$ is the total power of the mode as determined from the power envelope for radial modes \citep{Kjeldsen}, and $T$ is the effective length of the observational run. The heights of non-radial modes were then defined based on the heights of radial modes according to Eq.~\eqref{eqn:Ballot}, and taking into account the appropriate $V_\ell/V_0$ ratios given in table~1 of \citet{Kjeldsen}.

\item The background was parametrized as a linear function of frequency since it had previously been suppressed at low frequencies (high-pass cut at $\SI{280}{\mu Hz}$) to effectively remove slow variations.

\item The inclination angle between the direction of the stellar rotation axis and the line of sight was fixed at $31.1\,^{\circ}$, which is the inclination of the binary orbit and is consistent with the rotational modulation of the velocity curve. The rotational splitting was fixed at $\SI{0.7}{\mu Hz}$, which was chosen to match the observed value of $v \sin(i)\!=\!\SI{3.16}{km~s^{-1}}$ \citep{AllendePrieto2002}, given the known radius of the star of $\SI{2.05}{\Rsun}$ \citep{Kervella2004}.

\item We drew $\sim\num{800000}$ samples from the target distribution after a burn-in phase. We employed 12 parallel tempered chains.

\item We thus have a total of 46 free parameters, namely, 42 frequencies, 2 parameters for the linewidth and 2 parameters for the background.

\end{itemize}

\begin{figure}[h]
	\centering
	\includegraphics[width=0.5\textwidth]{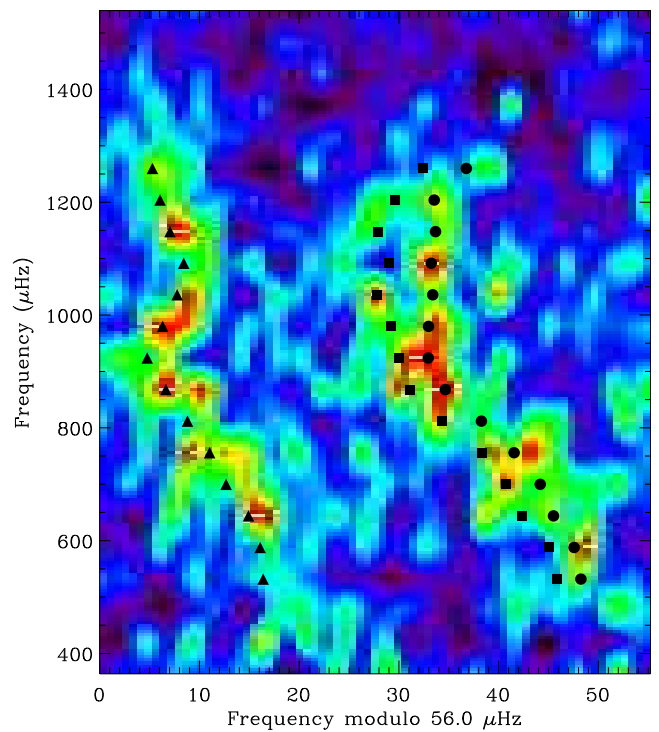}
	\caption{Power density spectrum of Procyon (smoothed to $\SI{2}{\mu Hz}$) in \'echelle format. The fitted frequencies for model $M_{\rm{A}}^{\ell=3}$ appear as overlaid filled symbols. Symbol shapes indicate mode degree: $\ell\!=\!0$ (circles), $\ell\!=\!1$ (triangles), $\ell\!=\!2$ (squares).}
	\label{fig:Procyon}
\end{figure}

Table~\ref{tab:ModProb} summarises the model selection calculations assuming equal prior probabilities for the models belonging to our discrete model space. Individual probabilities are assigned to models according to Eq.~\eqref{eqn:BayesTheorem3}. Similarly to \citet{ProcyonII}, Bayesian model comparison again statistically favours Scenario A over Scenario B. Furthermore, the presence of residual power due to $\ell\!=\!3$ modes is suggested. Computing Bayes' factor in favour of model $M_{\rm{A}}^{\ell=3}$ over model $M_{\rm{B}}^{\ell=3}$ gives a factor of approximately 9:1 or, equivalently, a logarithmic factor of 2.2, which classifies as `significant' on Jeffreys' scale. Figure~\ref{fig:Procyon} displays the power density spectrum of Procyon in \'echelle format with the fitted frequencies for model $M_{\rm{A}}^{\ell=3}$ overlaid. 
\begin{table}
	\centering
	\caption{Model probabilities.}
	\begin{tabular}{ccc}
		\toprule
		Model & $\ln p(D|{\rm{Model}},I)$ & Probability \\
		\midrule
		$M_{\rm{A}}$ & 2789.723 & 39.25\% \\
		$M_{\rm{A}}^{\ell=3}$ & 2790.046 & 54.23\% \\
		$M_{\rm{B}}$ & 2785.806 & 0.78\% \\
		$M_{\rm{B}}^{\ell=3}$ & 2787.801 & 5.74\% \\
		\bottomrule
	\end{tabular}
	\label{tab:ModProb}
\end{table}

\section{Summary and discussion}\label{sect:Conclusions}
In this paper, we have presented the basic theory and methods behind the extraction of parameters from the power spectra of solar-like stars.
In order to handle the ever rising quality and complexity of modern asteroseismic data, we have developed a tool (APT~MCMC) that enables us to constrain parameters associated with the subtlest features in the spectra. The algorithm has been extensively tested and performs extremely well, not only in the traditional case of extracting oscillation frequencies, but also when pushing the limit where traditional methods have difficulties, such as constraining linewidths, rotational splittings and stellar inclination angles.
In this work we have focused on data in the signal-to-noise regime of current asteroseismic measurements.
In the case of very high signal-to-noise ratios, other features in the power spectrum becomes important, such as mode asymmetries and rotational splittings dependent on $\ell$, arising from differential rotation with radius.
In future work these effects will be incorporated into the program and tested on solar data.

One disadvantage of the method is that it can be quite computationally intensive, both to implement and run, when compared to traditional MLE fits. This is however balanced by the much added information outputted from the fits, specifically in the probability distributions of each parameter, making it easy to obtain accurate, reliable and realistic error bars on the results -- a feature seriously missing from the traditional methods.
The parameter estimation also benefits enormously from the possibilities the Bayesian formalism provides with inclusion of prior information.
This not only allows control of the fit to, for example, not allow unphysical parameter combinations, but also include information into the fit that is better constrained by other measurements (as we saw in Sect.~\ref{sec:HD49933}).
Another powerful feature of the method lies in the parallel tempering, which not only keeps the fits from getting stuck in local maxima, but also provides an objective way of comparing different competing models, as it provides a way of calculating the global likelihood. This can for example be utilized in the familiar problem of ridge identification in solar-like stars (see Sect.~\ref{sec:Procyon}).

A thing to keep in mind is also that the APT~MCMC algorithm is completely general, in the sense that it could be applied to other problems without modification.
MCMC methods are being used in various branches of astrophysics: cosmology \citep{Liddle}, extra solar planets \citep{Gregory2005} and stellar model fitting \citep{Bazot}, but in fact the methods would be applicable in any problem including parameter estimation. And as computational power continues to grow, the downsides are quickly becoming insignificant.

What could to some extent also be seen as a disadvantage of these methods is that they can never be fully automated, in the sense that they will not be able to handle a large number of stars without human interaction.
The whole fundamental idea behind the Bayesian formalism is that it relies on ''wise`` human inputs on the priors and model setup that should not be done in an automated way. If nothing else, take this as a positive reassurance: You will, as an astrophysicist, never be obsolete to computers or monkeys with keyboards.

\begin{acknowledgements}
We would like to thank H. Kjeldsen, T. R. Bedding, C. Karoff, W. J. Chaplin, T. Appourchaux, R. A. Garc\'ia and O. Benomar for fruitful discussions and valuable comments.
We also thank the anonymous referee for comments that helped improve the paper.
TLC is supported by grant with reference number SFRH/BD/36240/2007 from FCT/MCTES, Portugal.
\end{acknowledgements}

\bibliography{15451}

\begin{appendix}
\section[Computing Elm and Vl]{Computing $\mathscr{E}_{\ell m}(i)$ and $V_\ell$}\label{sec:Visibilities}
The $\mathscr{E}_{\ell m}(i)$ factors are given below for $\ell \! \in \! [0,4]$, having used Eq.~\eqref{eqn:Elm}:
	\begin{equation}
	\begin{split}
		\mathscr{E}_{0,0}(i) &= 1 \, , \\
		\mathscr{E}_{1,0}(i) &= \cos^2 i \, , \\
		\mathscr{E}_{1,\pm1}(i) &= \tfrac{1}{2} \sin^2 i \, , \\
		\mathscr{E}_{2,0}(i) &= \tfrac{1}{4} (3\cos^2 i - 1)^2 \, , \\
		\mathscr{E}_{2,\pm1}(i) &= \tfrac{3}{8} \sin^2(2 i) \, , \\
		\mathscr{E}_{2,\pm2}(i) &= \tfrac{3}{8} \sin^4 i \, , \\
		\mathscr{E}_{3,0}(i) &= \tfrac{1}{64} (5\cos(3 i) + 3\cos i)^2 \, , \\
		\mathscr{E}_{3,\pm1}(i) &= \tfrac{3}{64} (5\cos(2 i) + 3)^2 \sin^2 i \, , \\
		\mathscr{E}_{3,\pm2}(i) &= \tfrac{15}{8} \cos^2 i \sin^4 i \, , \\
		\mathscr{E}_{3,\pm3}(i) &= \tfrac{5}{16} \sin^6 i \, , \\
		\mathscr{E}_{4,0}(i) &= \tfrac{1}{64} (35\cos^4 i - 30\cos^2 i + 3)^2 \, , \\
		\mathscr{E}_{4,\pm1}(i) &= \tfrac{5}{256} (\tfrac{7}{2} \sin(4 i) + \sin(2 i))^2 \, , \\
		\mathscr{E}_{4,\pm2}(i) &= \tfrac{5}{128} (7\cos(2 i) + 5)^2 \sin^4 i \, , \\
		\mathscr{E}_{4,\pm3}(i) &= \tfrac{35}{16} \cos^2 i \sin^6 i \, , \\
		\mathscr{E}_{4,\pm4}(i) &= \tfrac{35}{128} \sin^8 i \, .
	\end{split}
	\end{equation}
Notice that when the rotation axis is aligned with the line of sight ($i\!=\!0^\circ$), only the multiplet component with $m\!=\!0$ is visible, thus making inviable an inference of rotation.

The spatial response function for each $\ell$, $V_\ell$, representing the ratio of the observed amplitude to the actual amplitude, is given here for the five lowest degree modes \citep[][]{BeddingEtAl.1996}: 
\begin{equation}
		\begin{pmatrix} V_0 \\ V_1 \\ V_2 \\ V_3 \\ V_4 \end{pmatrix} = %
			\begin{pmatrix}
				1 & \tfrac{2}{3} & \tfrac{1}{2} & \tfrac{2}{5} \\
				\tfrac{2}{\sqrt{3}} & \tfrac{\sqrt{3}}{2} & \tfrac{2\sqrt{3}}{5} & \tfrac{1}{\sqrt{3}} \\
				\tfrac{\sqrt{5}}{4} & \tfrac{4}{3\sqrt{5}} & \tfrac{\sqrt{5}}{4} & \tfrac{8}{7\sqrt{5}} \\
				0 & \tfrac{\sqrt{7}}{12} & \tfrac{4}{5\sqrt{7}} & \tfrac{\sqrt{7}}{8} \\
				-\tfrac{1}{8} & 0 & \tfrac{3}{32} & \tfrac{16}{105}
			\end{pmatrix} \times %
			\begin{pmatrix}
				1-c & c-1 & c-1 \\
				c & 1-2c & -c \\
				0 & c & 1-c \\
				0 & 0 & c
			\end{pmatrix} \times %
			\begin{pmatrix}
			 	1 \\ u_2 \\ v_2
			\end{pmatrix} \, ,
	\end{equation}
where $u_2$ and $v_2$ are wavelength-dependent classical limb-darkening coefficients \citep[][]{Allen1973} and $c$ is a parameter defining the observational method. This matrix product can be used for velocity measurements by setting $c\!=\!1$ and for intensity measurements by setting $c\!=\!0$.
\end{appendix}
\end{document}